\documentclass[10pt,preprint]{aastex}

\newcommand{\av}{$A_V$}

\newcommand{\etal}{et~al.}

\newcommand{\ks}{$K_{\rm s}$}

\newcommand{\mdot}{$\dot{M}$}
\newcommand{\msun}{M$_{\sun}$}

\newcommand{\teff}{$T_{\rm eff}$}

\newcommand{\mum}{$\mu$m}

\begin{document}

\title{The Evolution of Circumstellar Disks Surrounding Intermediate
Mass Stars: IC 1805}

\slugcomment{Version from \today}

\author{
S.\ C.\ Wolff\altaffilmark{1},
S.\ E.\ Strom\altaffilmark{1},
L.\ M.\ Rebull\altaffilmark{2}
}

\altaffiltext{1}{NOAO, 950 N. Cherry Ave, Tucson, AZ (swolff@noao.edu)}
\altaffiltext{2}{Spitzer Science Center/Caltech, Pasadena, CA  91125}

\begin{abstract}

We report the results of a study of the intermediate and high mass
stars in the young, rich star-forming complex IC 1805, based on a
combination of optical, near-infrared, and  mid-infrared photometry,
and classification spectra. These data provide the basis for
characterizing the masses and ages for stars more massive than $\sim2$
\msun\ and enable a study of the frequency and character of
circumstellar disks associated with intermediate- and high-mass
stars.  Optically thick accretion disks among stars with masses  $2 <
M/M_{\odot} <4$ are rare ($\sim$2\% of members) and absent among more
massive stars. A larger fraction ($\sim$10\%) of stars with masses $2
< M/M_{\odot} < 4$ appear to be surrounded by disks that have evolved
from the initial optically thick accretion phase.   We identify four
classes of such disks.  These classes are based on spectral energy
distributions (SEDs) of excess emsission above photospheric levels: 
disks that are (1) optically thin based on the magnitude of the
observed excess emission from 2 to 24 \mum;  (2) optically thin in
their inner regions ($r<$ 20 AU) and optically thick in their outer
regions; (3) exhibit empty inner regions ($r <$ 10 AU) and optically
thin emission in their outer regions; and (4) exhibit empty inner
regions and optically thick outer regions.  We discuss, and assess the
merits and liabilities of, proposed explanations for disks exhibiting
these SED types and suggest additional observations that would test
these proposals.

\end{abstract}


\section{Introduction}
\label{sec:intro}

The advent of sensitive ground- and space- based infrared (IR)
instrumentation has provided  astronomers with the tools to determine
the evolutionary history of the circumstellar  disks that appear to
surround stars of all masses at birth. Spectral energy distributions 
(SEDs) that exhibit excess IR emission above photospheric levels
provide the  basis for detecting such disks and inferring the radial
and vertical distribution of disk  material. Early studies focused on
establishing the timescales over which disks survive  as optically
thick accretion disks (e.g., Strom \etal\ 1989). These initial results
placed an  important constraint on the timescales over which disks are
likely to form planets. Later,  astronomers began to focus on changes
in the radial distribution of small dust grains  (Strom \etal\ 1989;
Skrutskie \etal\ 1990) in order to search for disks that have begun
to  `transition' from their initial, optically thick accretion phase
to more advanced  evolutionary states. Such studies provided the
first hints of changes wrought by  processes such as photoevaporation,
planetesimal formation, and giant planet formation  (see, for example,
the original discussion in Skrutskie \etal\ 1990; Clarke, Gendrin, \& 
Sotomayor 2001). From studies of large samples of low-mass ($M <$ 1
\msun) stars, it is  now generally accepted that (a) the fraction of
stars surrounded by optically thick  accretion disks decreases from
$\sim$80-90\% among the youngest observable stellar  populations, to
$\sim$50\% at 3 Myr and $\sim$10\% at 5 Myr (Haisch, Lada, \& Lada
2001); (b) a  modest fraction of disks ($\sim$5-15\%) in clusters
ranging in age from 1-5 Myr appear to  exhibit IR SEDs that suggest
significant evolution from an initial, optically thick state.  
Examples include disks with optically thin inner holes and optically
thick outer disks, disks that show evidence of grain settling and
possible grain growth, and disks that have SEDs consistent with emission
from optically thin dust or gas (see Currie \etal\ 2009 or Cieza
\etal\ 2010 for a recent review).

While much attention has been devoted to understanding disk evolution
around solar-like stars, relatively little work has focused on the
early stages of disk evolution among  higher mass objects, largely
because robust samples of nearby, young intermediate  mass objects are
not available. Early work (Strom 1972; Strom \etal\ 1972; Hillenbrand 
\etal\ 1993) as well as more recent work (Dahm \& Hillenbrand 2007;
Currie \& Kenyon  2009) suggested that the fraction of massive stars
surrounded by optically thick  accretion disks at a given age is
considerably smaller than the fraction of such disks  found among
solar-like stars.  

We report here the results of a study of IC 1805, a young, rich
cluster located at a distance of 2350 pc in the  molecular cloud
associated with W4 (Vasilevskis, Sanders, \& van Altena 1965;  Sagar
\etal\ 1988). Our goal is to take advantage of the large ($>$500
stars) population of B and  A stars in this region to quantify the
fraction of intermediate mass stars surrounded by  optically thick
accretion disks and to search for and understand the nature of disks 
transitioning from this phase.

Our study relies on a combination of optical, near-IR, and Spitzer
Space Telescope (Werner \etal\ 2004) mid-IR photometry  combined with
classification spectra obtained for a large sample of IC 1805
members.  We make use of the photometry and classification spectra to
(a) locate stars in an  observational Hertzpring-Russell (HR) diagram
and to determine ages and masses for members; and (b) to  derive
reddening-corrected spectral energy distributions, which enable us to
assess  both the fraction of stars surrounded by optically thick
accretion disks and the number  and character of various types of disks in
more advanced evolutionary states.

We first discuss our sample and present the photometric data and their
uncertainties (\S\ref{sec:obs});  describe our methods for determining
likely members of IC 1805 and their ages and masses
(\S\ref{sec:properties}); and report our results for a sample of 63
stars with IR excesses and  masses $M >$ 2 \msun\ drawn from a sample
of 548 likely members of this cluster (\S\ref{sec:irx}).  We use
reddening-corrected SEDs  to identify stars surrounded by optically
thick accretion disks, as well as four classes of  objects whose SEDs
suggest the presence of disks in different  physical states. These
states presumably represent alternative paths, or ``next steps'' for
disks as they evolve from initial optically thick accretion disks to
more advanced evolutionary states.  In \S\ref{sec:discussion}, we discuss
the possible physical mechanisms that lead  to these four SED classes
(and presumably different evolutionary states), assess the merits and 
liabilities of each of the 
proposed mechanisms, and suggest observational tests aimed at sorting
among these  possibilities.

\section{Observations}
\label{sec:obs}

\subsection{Target Selection and Optical/NIR Photometry}

The primary goal of the current study is to determine the disk
properties of intermediate  mass stars in IC 1805.  We observed this
region with Spitzer using the Infrared Array Camera (IRAC; Fazio
\etal\ 2004a) and the Multiband Imaging Photometer for Spitzer (MIPS;
Rieke \etal\ 2004).  We used the literature to identify candidate
members; the $UBV$ photometric study by Massey \etal\ (1995a) provides
a  list of 1023 optically-selected candidate members of IC 1805 down
to a mass of about 2 \msun.  Most of these objects are within the
region covered by our IRAC observations; just 220 of these stars lie
outside the IRAC map.  We matched the optical sources to sources in
the Two-Micron All-Sky Survey (2MASS; Skrutskie \etal\ 2006) to obtain
internally consistent positions and near-IR photometry. Two of the
optical sources (Massey \etal\ Nos.\ 292 and 461) were identified with
a single 2MASS source.   We have arbitrarily assigned the NIR
measurements to No.\ 461, which is the brighter of  the two.  There
are therefore 802 stars that are potential members of IC 1805 that
have  been observed with IRAC.  The MIPS map covers a region that is 
somewhat larger than that covered by the IRAC map. Consequently, we
have Spitzer  detections or upper limits for at least one wavelength
for 974 of the stars included in the  study by Massey \etal\ (1995a).

\subsection{IRAC}

Spitzer/IRAC observes at 3.6, 4.5, 5.8, and 8 \mum.  In order to
extract magnitudes for the 802 IC 1805 stars observed in the IRAC
bands, we started with the Spitzer Science Center (SSC)
pipeline-produced basic calibrated data (BCDs), version S14.4, for the
IRAC data from our program, 20052, AORKEY 13846016.  These
observations were 12 second high dynamic range (HDR) observations
(meaning short and long exposures are taken at each pointing), in a
9$\times$9 square map, with 5 medium dithers per pointing for a total
integration time of $\sim$60 seconds per pointing.  We ran the IRAC
Artifact Mitigation code written by S.\  Carey and available on the
SSC website. We constructed a mosaic from the corrected  BCDs using
the SSC mosaicking and point-source extraction (MOPEX) software 
(Makovoz \& Marleau 2005), with a pixel scale of 1.22$\arcsec$
px$^{-1}$, very close to the native pixel scale.  Our final map covers
$\sim$0.5 square degrees, centered on 02:32:42, +61:27:00 (see
Fig.~\ref{fig:backgroundism1} below).  Note that the 3.6 and 5.8 \mum\
channels (ch.\ 1 \& 3) share a field of view which is offset from the
field of view shared by the 4.5 and 8 \mum\ channels (ch.\ 2 \& 4),
and thus the maps for 3.6/5.8 cover the same total area but a region
of sky offset by $\sim$5$\arcmin$ northwest from the 4.5/8 maps.

Using an IDL photometry routine, we performed aperture  photometry on
the known target positions in the combined mosaic for the short and
long  exposures separately, using a 3-pixel aperture and a sky annulus
of 3-7 pixels. The  (multiplicative) aperture corrections we used
follow the values given in the IRAC Data  Handbook: 1.124, 1.127,
1.143, and 1.234 for IRAC channels 1, 2, 3, and 4,  respectively. For
stars brighter than magnitude 9.5, 9.0, 8.0, and 7.0 for IRAC-1, 2,
3,  and 4, respectively, we took the flux densities from the short
rather than the long  exposure. We took the errors returned by the IDL
photometry routine, which are  statistical in nature, and added them
in quadrature to a 5\% flux density error floor. With this floor, 91\%
of the IRAC-1 (3.6 \mum) sources have errors less than 0.06  mag, 80\%
of the IRAC-2 (4.5 \mum) sources have errors less than 0.06 mag, 95\%
of the IRAC-3 (5.8 \mum)  sources have errors less than 0.1 mag, and
60\% of the IRAC-4 (8 \mum) sources have errors  less than 0.15 mag.

We compared the flux densities as obtained from a 3-pixel aperture and
a sky annulus  of 3-7 pixels (with the aperture corrections as listed
above) to that obtained from a 2  pixel aperture and a sky annulus of
2-6 pixels (with the appropriate aperture corrections  as listed on
the SSC website). In most cases, the flux densities agreed to well
within the  5\% flux density error floor. In all of the remaining
cases, the errors were within the 0.3  magnitudes estimated to be the
uncertainty due to variation in local reddening as  discussed in
\S\ref{sec:finalirx}.  

The photometric observations for this optically-selected sample are
reported in  Table~\ref{tab:spitzerfluxes}.  Column 1 lists the
optical number from Massey \etal\ (1995a);  columns 2 and 3 give the
RA and Dec; column 4 gives the 2MASS name; columns 5-7  list the $UBV$
magnitudes from Massey \etal; columns 8-10 give the $JHK_s$ values
from  2MASS; columns 11-14 give the IRAC magnitudes and errors (or
limit); and column 15 gives the MIPS magnitude and errors (or limit)
-- see next section.  The total numbers of objects from the
optically-selected sample detected (or for which we have limits) for
each band are listed in Table~\ref{tab:totaldet}.

\begin{deluxetable}{lrrrrrrrrrrrrrrr}
\tablecaption{Spitzer measurements for sample of previously identified
IC 1805 members\tablenotemark{a}\label{tab:spitzerfluxes}}
\tabletypesize{\tiny}
\rotate
\tablewidth{0pt}
\tablehead{
\colhead{(1)} & \colhead{(2)} & \colhead{(3)} &
\colhead{(4)} & \colhead{(5)} &
\colhead{(6)} & \colhead{(7)} & 
\colhead{(8)} & \colhead{(9)} & 
\colhead{(10)} & \colhead{(11)} & 
\colhead{(12)} & \colhead{(13)} & 
\colhead{(14)} & \colhead{(15)} \\
\colhead{Opt.} & \colhead{RA} & \colhead{Dec} &
\colhead{2MASS name} & \colhead{$U$} &
\colhead{$B$} & \colhead{$V$} & 
\colhead{$J$} & \colhead{$H$} & 
\colhead{$K_s$} & \colhead{$[3.6]$} & 
\colhead{$[4.5]$} & \colhead{$[5.8]$} & 
\colhead{$[8]$} & \colhead{$[24]$} \\ 
\colhead{num.} & \colhead{(J2000; deg)} & \colhead{(J2000; deg)} &
\colhead{ } & \colhead{(mag)} &
\colhead{(mag)} & \colhead{(mag)} & 
\colhead{(mag)} & \colhead{(mag)} & 
\colhead{(mag)} & \colhead{(mag)} & 
\colhead{(mag)} & \colhead{(mag)} & 
\colhead{(mag)} & \colhead{(mag)}}
\startdata
       1&   38.620958&   61.421306&  02342903+6125167&       16.50&       16.19&       15.17&    13.23$\pm$  0.03&    12.98$\pm$  0.04&    12.83$\pm$  0.04&    12.70$\pm$  0.06&    12.71$\pm$  0.06&    12.47$\pm$  0.08&    12.00$\pm$  0.12&            $>$   7.15\\
       2&   38.617125&   61.343194&  02342811+6120355&       16.90&       16.20&       15.20&    13.07$\pm$  0.02&    12.65$\pm$  0.03&    12.52$\pm$  0.03&    12.49$\pm$  0.06&    12.47$\pm$  0.06&    12.41$\pm$  0.07&    12.16$\pm$  0.10&            $>$  10.15\\
       3&   38.603875&   61.556361&  02342493+6133229&       16.09&       15.63&       14.87&    13.05$\pm$  0.02&    12.76$\pm$  0.03&    12.56$\pm$  0.03&    12.48$\pm$  0.06&    12.46$\pm$  0.06&    12.45$\pm$  0.06&    12.56$\pm$  0.07&            $>$   8.98\\
       4&   38.602333&   61.548278&  02342456+6132538&       14.69&       14.73&       14.01&    12.07$\pm$  0.04&    11.87$\pm$  0.03&    11.70$\pm$  0.04&    11.35$\pm$  0.06&    11.20$\pm$  0.06&    11.07$\pm$  0.06&    10.81$\pm$  0.06&     9.94$\pm$  0.13\\
       5&   38.595042&   61.421250&  02342281+6125165&       16.28&       15.72&       14.70&    12.66$\pm$  0.11&    12.13$\pm$  0.03&    12.00$\pm$  0.03&    11.97$\pm$  0.06&    12.00$\pm$  0.06&    11.91$\pm$  0.07&    11.86$\pm$  0.14&            $>$   9.39\\
\enddata
\tablenotetext{a}{Table will be presented in its entirety in the
electronic version of the Journal. A portion is presented here as a
guide to its form and content.}
\end{deluxetable}

\begin{deluxetable}{lllcccrrr}
\tablecaption{Total Detected or Constrained Objects from the Optically-Selected
Sample\label{tab:totaldet}}
\tablewidth{0pt}
\tablehead{
\colhead{band} & \colhead{total detected} & 
\colhead{total with limits} }
\startdata
IRAC 1 (3.6 \mum) & 678 & 7\\
IRAC 2 (4.5 \mum) & 712 & 4 \\
IRAC 3 (5.8 \mum) & 685 & 0 \\
IRAC 4 (8 \mum)   & 708 & 0 \\
MIPS 1 (24 \mum)  & 48 & 869 \\
\enddata
\end{deluxetable}

\subsection{MIPS}

\begin{figure*}[tbp]
\epsscale{1.0}
\plotone{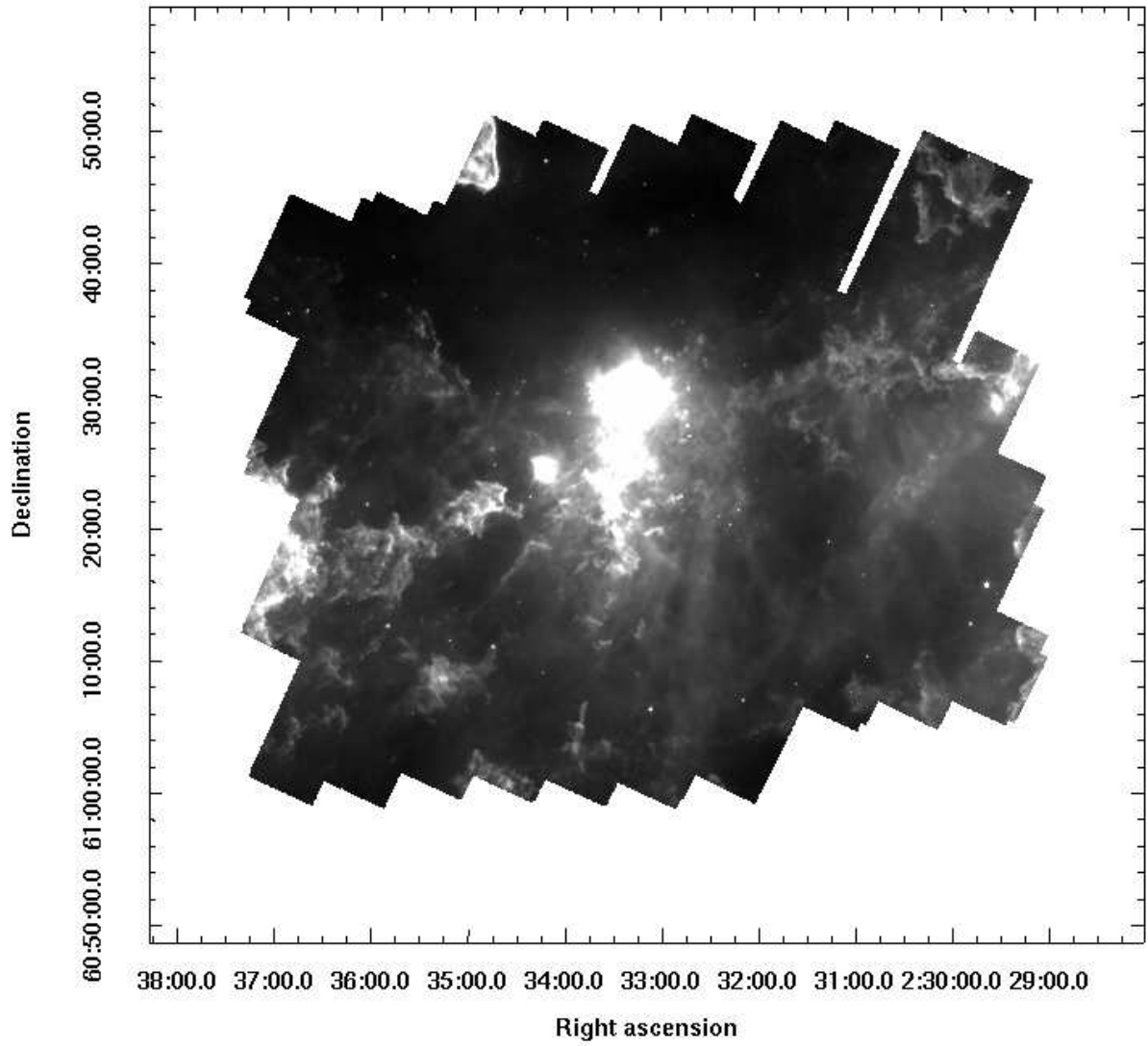}
\caption{Mosaic of 24 \mum\ map covering the IC 1805 region. North is
up.   }
\label{fig:24ummap}
\end{figure*}

Spitzer/MIPS observations were made at 24, 70, and 160 \mum.  As for
IRAC, we started with the SSC pipeline-produced BCDs, downloading data
in this region, both from our program (20052) and another, 3234 (PI
J.\ Greaves). The data from 20052 (AORKEYs 13846272, 13846528, 
13846784) were processed under S16.0.1, but the data from 3234
(AORKEYs  10498048 and 10498304) were processed under S14.4. (For a
description of the  pipeline, see Gordon \etal\ 2005.) The only
material difference in these pipeline versions is in the treatment of
very bright sources, and none of our targets crosses this threshold. 
The observations from program 20052 were a 5$\times$5 raster map of
3-second, 7-cycle small-field photometry-mode observations, resulting
in an integration time of $\sim$312 seconds per position.  The
observations from program 3234 were fast scan maps with 302$\arcsec$
cross-scan steps (95\% of detector width), for a total of $\sim$15
seconds per position.  We combined all of the BCDs using MOPEX into
one $\sim0.7$ deg$^2$ mosaic centered on the region of interest
(roughly 02:33:10, +61:26:20), with a pixel scale of 2.5$\arcsec$
px$^{-1}$, close to the native pixel scale. In most of this region,
the total integration time is $\sim$327 seconds, but it varies
according to which (and how many) BCDs were included at a given
location.  Figure~\ref{fig:24ummap} shows the complete 24 \mum\ mosaic
in the IC 1805 region.  Sensitivity is a strong function of location
not only because of the number of BCDs included at any given position,
but also because of the wide variation in sky brightness. 

We extracted sources from our 24 \mum\ mosaics using the APEX-1-frame
portion of  MOPEX, with point response function (PRF)-fitting
photometry of the image mosaics.  For six bright sources, aperture
photometry was found to be a better measure of the  total flux density
from the object; for these objects, we used a 13$^{\prime\prime}$ 
aperture, with a 20-32$^{\prime\prime}$ sky annulus, and an aperture
correction of 1.17  as tabulated on the SSC website. For errors on
these values, we derived the errors from  the signal-to-noise as
returned by APEX. These are statistical uncertainties; the  systematic
uncertainty in the zero-point of the conversion from instrumental
units to  calibrated flux density units is estimated to be 4\%
(Engelbracht \etal\ 2007). So, for the  errors reported in the table
above, we added 4\% in quadrature to the errors derived  from the APEX
results.

Most of our optically-identified objects are not detected at 24 \mum.
In order to obtain  upper limits, we placed an aperture and annulus at
the location of the corresponding  optical source and took the
absolute value of the difference between the aperture and  annulus
flux density. We tested different apertures, and found that the limits
obtained for  the same aperture as used for the bright sources were
unacceptable because they often  included faint background sources. As
a result, we chose to use a small aperture of  3.5$^{\prime\prime}$,
an annulus of 20-32$^{\prime\prime}$, and an aperture correction  of
2.57 (again, as tabulated on the SSC website) to obtain these upper
limits. We then  multiplied these values by 5 to obtain the 5$\sigma$
upper limit values reported in the last column of
Table~\ref{tab:spitzerfluxes}. We tested the viability of these
automatically-determined values by spot-checking several undetected
targets in a variety of  background regions and found good agreement.

No viable data were obtained at 70 or 160 \mum\ for our targets.

\section{Properties of IC 1805}
\label{sec:properties}
\subsection{Membership Criteria}

In order to analyze the disk properties of intermediate mass stars in
IC 1805 as a  function of stellar mass and age, we must first separate
cluster members from  contaminating field stars.  The photometric and
spectroscopic data provide three criteria  that can be used to select
likely members:  1) reddening consistent with that measured  for the
definite early type (B2.5V and earlier) members selected by Massey
\etal\  (1995a); 2) the presence of the interstellar feature
$\lambda$4430 in classification dispersion  spectra, which is absent
in the spectra of foreground stars; and 3) position in the HR 
diagram.  

To assist with the determination of reddening, we obtained spectra of
a sample of 229 of  the 802 candidates. All but approximately 20 of
these stars were selected at random;  the exceptions were those
objects which appeared to have obvious and strong IR  excesses based
on our preliminary reductions of the IRAC data. The observations were 
obtained with the Hydra multi-object fiber spectrograph at the WIYN
telescope on October 30 to November 2, 2007.  The  spectra covered the
wavelength range 3600 to 5300 \AA\ at a resolution of 1.3 \AA\
px$^{-1}$ and a typical signal-to-noise ratio (SNR) of 30-100. 
Spectral classification was effected  by comparing the  stars in IC
1805 with well-studied stars in the Pleiades, which were used as
standards (Crawford \& Perry 1976).  We estimate that the maximum
likely  error in type is $\pm$ two subtypes for stars A0 and later. 
The uncertainty for the early B-type  stars could be as much as 5
subtypes because of the lack of suitable  standards earlier than B7 in
the Pleiades.  For these hotter stars,  however, we can derive the
reddening from $UBV$ photometry alone (see below).

In their study, Massey \etal\ (1995a) obtained spectra for 38
early-type stars (B2.5V to  O4) and found that the reddening for
cluster members fell in the range 0.68$<E(B- V)<$1.29 mag.  For early
type stars (those with $Q = (U-B) - 0.72 \times (B-V) < -0.4$), we
can  use the $Q$ method to determine the reddening (Massey \etal\
1995b).  Basically, this  procedure involves using the $Q$-index,
which is independent of reddening, to derive the  intrinsic color
$(B-V)_0$ and then comparing the intrinsic and observed colors to
determine  the reddening.  For the cooler stars ($Q > -0.4$) this
method does not yield a unique  solution for the intrinsic color, and
classification spectra are required.  We have also  correlated the
measured reddening with the presence or absence of the interstellar 
feature $\lambda$4430.  We find that stars with a detectable
$\lambda$4430 band (equivalent width $>$ 0.5 \AA)   have reddening in
the range $0.5 < E(B-V) < 1.30$, a range only slightly broader than
that  estimated by Massey \etal\ (1995a).  By extending the lower
limit for membership to  $E(B-V)$ of 0.5, we include all but four of
the 63 stars with IR excesses (see \S\ref{sec:irx} for the
determination of IR excesses);  stars with IR excesses are likely also
to be members of  IC 1805.

For the purposes of this study, therefore, we will use reddening of
$0.5<E(B-V)<1.30$  coupled with a location in the HR diagram that is
consistent with membership in IC 1805  as our two primary membership
criteria.  We have used the $Q$ method to estimate  reddening for
stars with $Q< -0.4$ and $U < 15$.  The $U$ photometry becomes
increasingly  less accurate for stars fainter than $U$=15 (Massey
\etal\ 1995a), and for these fainter  stars and for all stars with $Q>
-0.4$ we have estimated the reddening from the spectral  types.  

The 229 stars that meet our criteria for membership in IC 1805 are
listed in Table~\ref{tab:members}.  The  first column gives the star
number assigned by Massey \etal\ (1995a).  (The members that have IR
excesses are marked with an asterisk.) The second column gives the
spectral type from the current study, and the third column lists the 
equivalent width of $\lambda$4430 if it was detected.  The fourth
column gives the reddening  derived from the $Q$ method for those
stars with $Q<-0.4$, and the fifth column gives the reddening derived
from the spectral type.  The sixth column gives the adopted reddening
$E(B-V)$, which as noted  above was derived from $Q$ for the hot stars
for which this method is valid or from  spectral types for the cooler
stars.  The remaining columns give the reddening corrected  values of
$B_0$, $V_0$, and $(B-V)_0$ that will be used below to construct
color-magnitude and  color-color plots.

An additional 27 stars have IR excesses but no independent measurement
of  reddening, and four more stars with excesses have reddening based
on spectral types  in the range $0.30 < E(B-V) < 0.42$.  These 31
stars do fall in a reasonable place in an  HR diagram after correction
for the mean reddening of IC 1805 (Massey \etal\ 1995a),  and we will
assume that all 31 are members of IC 1805 by virtue of these
excesses. 

Of the remaining stars, 151 can be rejected on the basis of their
reddening or positions  in an HR diagram.  Table~\ref{tab:rejects}
summarizes the number of stars rejected and the reasons  for
rejection.

\begin{deluxetable}{lllcccrrr}
\tablecaption{Likely Members of IC 1805\tablenotemark{a}\label{tab:members}}
\tabletypesize{\tiny}
\rotate
\tablewidth{0pt}
\tablehead{
\colhead{(1)} & \colhead{(2)} & \colhead{(3)} & \colhead{(4)} &
\colhead{(5)} & \colhead{(6)} & \colhead{(7)} & \colhead{(8)} & \colhead{(9)}\\
\colhead{opt.\ no.\tablenotemark{b}} & \colhead{spectral} & 
\colhead{EQW($\lambda$4430)} &
\colhead{$E(B-V)$ from}  &
\colhead{$E(B-V)$ from} & \colhead{adopted } & \colhead{$B_0$} 
& \colhead{$V_0$} & \colhead{$(B-V)_0$} \\
 & \colhead{type} & \colhead{(\AA)} & \colhead{$Q$ method} 
 & \colhead{spec.\ type} & \colhead{$E(B-V)$} & \colhead{(mag)} &\colhead{(mag)} 
 &\colhead{(mag)} \\
  & & & \colhead{(mag)}& \colhead{(mag)}& \colhead{(mag)}}
\startdata
      3   & B9   &             1.3  &      \nodata   &       0.86   &    0.86  &    12.10  &    12.20  &    -0.10\\ 
      4*  & B8   &             1.3  &      \nodata   &       0.82   &    0.82  &    11.36  &    11.47  &    -0.10\\ 
      6*  & B8   &               1  &      \nodata   &       0.80   &    0.80  &    11.76  &    11.86  &    -0.10\\ 
      8   & A3   &             1.4  &      \nodata   &       0.63   &    0.63  &    12.17  &    12.10  &     0.08\\ 
     13*  & B7   &             1.5  &     0.81  &       0.64   &    0.81  &     8.40  &     8.68  &    -0.28\\ 
     15*  & B7   &            1.25  &       \nodata  &       0.82   &    0.82  &    11.84  &    11.95  &    -0.10\\ 
     17   & A2   &               1  &       \nodata  &       0.59   &    0.59  &    12.42  &    12.37  &     0.05\\ 
     18*  & A0   &             1.2  &     1.00  &       0.88   &    0.88  &    11.47  &    11.52  &    -0.05\\ 
     19   & F1   &             \nodata   &       \nodata  &       0.79   &    0.79  &    12.31  &    11.99  &     0.32\\ 
     21   & A0   &            1.35  &       \nodata  &       0.53   &    0.53  &    13.56  &    13.59  &    -0.02\\ 
     25   & \nodata   &             \nodata   &     0.80  &       \nodata     &    0.80  &     6.60  &     6.91  &    -0.31\\ 
     31   & \nodata   &             \nodata   &     1.03  &       \nodata     &    1.03  &    10.26  &    10.51  &    -0.24\\ 
     36   & A0   &             1.3  &       \nodata  &       0.73   &    0.73  &    12.69  &    12.72  &    -0.02\\ 
     40   & \nodata   &             \nodata   &     0.78  &       \nodata     &    0.78  &    11.02  &    11.19  &    -0.16\\ 
     41   & \nodata   &             \nodata   &     1.07  &       \nodata     &    1.07  &    10.01  &    10.17  &    -0.16\\ 
     43   & A0   &             \nodata   &     0.70  &       0.56   &    0.70  &    11.32  &    11.48  &    -0.17\\ 
     45   & F0   &             \nodata   &       \nodata  &       0.65   &    0.65  &    13.09  &    12.80  &     0.30\\ 
     46   & \nodata   &             \nodata   &     0.81  &       \nodata     &    0.81  &    10.46  &    10.67  &    -0.21\\ 
     49   & \nodata   &             \nodata   &     0.84  &       \nodata     &    0.84  &     8.22  &     8.49  &    -0.27\\ 
     50   & \nodata   &             \nodata   &     0.88  &       \nodata     &    0.88  &     9.80  &     9.95  &    -0.15\\ 
     55   & F6   &             \nodata   &       \nodata  &       0.56   &    0.56  &    13.42  &    12.96  &     0.46\\ 
     57   & F2   &             \nodata   &       \nodata  &       0.53   &    0.53  &    13.24  &    12.90  &     0.35\\ 
     59   & B5   &             1.1  &     0.72  &       0.66   &    0.72  &    11.60  &    11.76  &    -0.16\\ 
     63   & \nodata   &             \nodata   &     0.77  &       \nodata     &    0.77  &    11.39  &    11.55  &    -0.16\\ 
     65*  & A5   &      2.0:        &     \nodata    &       0.73   &    0.73  &    12.93  &    12.79  &     0.15\\ 
     67   & F0   &             \nodata   &       \nodata  &       0.66   &    0.66  &    12.76  &    12.46  &     0.30\\ 
     71   & A5   &             \nodata   &       \nodata  &       1.12   &    1.12  &    11.16  &    11.27  &    -0.10\\ 
     72   & \nodata   &             \nodata   &     1.24  &       \nodata     &    1.24  &     9.39  &     9.79  &    -0.41\\ 
     74   & \nodata   &             \nodata   &     0.75  &       \nodata     &    0.75  &     5.76  &     6.08  &    -0.32\\ 
     75   & \nodata   &             \nodata   &     0.86  &       \nodata     &    0.86  &    10.06  &    10.31  &    -0.24\\ 
     77   & \nodata   &             \nodata   &     0.92  &       \nodata     &    0.92  &    10.72  &    10.94  &    -0.22\\ 
     78   & F0   &             \nodata   &       \nodata  &       0.64   &    0.64  &    13.18  &    12.89  &     0.30\\ 
     79   & \nodata   &             \nodata   &      0.75 &       \nodata     &    0.75  &    10.10  &    10.32  &    -0.22\\ 
     85*  & B6   &             1.1  &     \nodata    &       0.75   &    0.75  &    12.18  &    12.29  &    -0.10\\ 
     86*  & A0   &             0.9  &     \nodata    &       0.91   &    0.91  &    12.17  &    12.20  &    -0.02\\ 
     88   & \nodata   &             \nodata   &     0.88  &       \nodata     &    0.88  &     8.53  &     8.83  &    -0.29\\ 
     90   & \nodata   &             \nodata   &     0.85  &       \nodata     &    0.85  &     8.22  &     8.51  &    -0.29\\ 
     92   & \nodata   &             \nodata   &     0.81  &       \nodata     &    0.81  &    10.64  &    10.84  &    -0.20\\ 
     93*  & A4   &             1.3  &     \nodata    &       0.84   &    0.84  &    12.31  &    12.19  &     0.13\\ 
     94   & \nodata   &             \nodata   &     0.88  &       \nodata     &    0.88  &     9.98  &    10.21  &    -0.23\\ 
     95   & G0   &             \nodata   &       \nodata  &       0.58   &    0.58  &    12.68  &    12.10  &     0.58\\ 
     99   & \nodata   &             \nodata   &     0.86  &       \nodata     &    0.86  &    10.14  &    10.36  &    -0.22\\ 
    100   & \nodata   &             \nodata   &     0.80  &       \nodata     &    0.80  &     8.84  &     9.09  &    -0.26\\ 
    102   & \nodata   &             \nodata   &     0.74  &       \nodata     &    0.74  &    10.91  &    11.08  &    -0.17\\ 
    106   & \nodata   &             \nodata   &     0.92  &       \nodata     &    0.92  &     8.63  &     8.89  &    -0.26\\ 
    107*  & A2   &             1.4  &     \nodata    &       0.75   &    0.75  &    11.42  &    11.38  &     0.05\\ 
    108   & \nodata   &             \nodata   &     0.68  &       \nodata     &    0.68  &    10.01  &    10.24  &    -0.23\\ 
    109   & \nodata   &             \nodata   &     0.81  &       \nodata     &    0.81  &     9.25  &     9.49  &    -0.24\\ 
    110   & B4   &             1.1  &     0.77  &       0.66   &    0.77  &    10.19  &    10.41  &    -0.21\\ 
    111   & B4   &             1.5  &     \nodata    &       1.01   &    1.01  &    11.75  &    11.86  &    -0.10\\ 
    112   & B5   &             1.1  &     0.96  &       0.88   &    0.96  &    10.76  &    10.95  &    -0.18\\ 
    113   & F8   &             \nodata   &       \nodata  &       0.60   &    0.60  &    12.91  &    12.39  &     0.52\\ 
    114   & \nodata   &             \nodata   &     0.95  &       \nodata     &    0.95  &     9.03  &     9.28  &    -0.25\\ 
    115   & \nodata   &             \nodata   &     0.74  &       \nodata     &    0.74  &     9.96  &    10.18  &    -0.22\\ 
    121   & B5   &             1.1  &     \nodata    &       0.92   &    0.92  &    11.59  &    11.70  &    -0.10\\ 
    124   & \nodata   &             \nodata   &     1.01  &       \nodata     &    1.01  &     4.62  &     4.94  &    -0.31\\ 
    125   & \nodata   &             \nodata   &     1.03  &       \nodata     &    1.03  &    10.26  &    10.51  &    -0.25\\ 
    126   & \nodata   &             \nodata   &     1.22  &       \nodata     &    1.22  &     8.72  &     8.98  &    -0.26\\ 
    128*  & A5   &             1.4  &     \nodata    &       0.93   &    0.93  &    12.07  &    11.93  &     0.14\\ 
    129   & A1   &             1.2  &     \nodata    &       0.79   &    0.79  &    12.66  &    12.65  &     0.01\\ 
    130   & \nodata   &             \nodata   &     0.86  &       \nodata     &    0.86  &    10.54  &    10.76  &    -0.23\\ 
    131   & \nodata   &             \nodata   &     0.87  &       \nodata     &    0.87  &     9.06  &     9.34  &    -0.28\\ 
    136   & \nodata   &             \nodata   &     0.83  &       \nodata     &    0.83  &     7.79  &     8.08  &    -0.29\\ 
    137*  & A2   &             0.8  &     \nodata    &       0.78   &    0.78  &    11.73  &    11.67  &     0.06\\ 
    138   & \nodata   &             \nodata   &     0.95  &       \nodata     &    0.95  &     9.73  &     9.96  &    -0.23\\ 
    139*  & A6   &             0.9  &     0.96  &       0.87   &    0.96  &    10.75  &    10.94  &    -0.20\\ 
    142   & \nodata   &             \nodata   &     0.74  &       \nodata     &    0.74  &     8.67  &     8.92  &    -0.25\\ 
    143*  & F4   &             \nodata   &       \nodata  &       0.51   &    0.51  &    13.81  &    13.41  &     0.40\\ 
    144*  & \nodata   &             \nodata   &     0.81  &       \nodata     &    0.81  &     5.09  &     5.40  &    -0.31\\ 
    145   & \nodata   &             \nodata   &     0.90  &       \nodata     &    0.90  &    10.34  &    10.53  &    -0.19\\ 
    146   & \nodata   &             \nodata   &     0.77  &       \nodata     &    0.77  &    10.90  &    11.08  &    -0.19\\ 
    148   & \nodata   &             \nodata   &     0.79  &       \nodata     &    0.79  &     8.65  &     8.94  &    -0.28\\ 
    153   & \nodata   &             \nodata   &     0.79  &       \nodata     &    0.79  &    10.35  &    10.53  &    -0.18\\ 
    154   & A3   &             1.5  &     \nodata    &       0.82   &    0.82  &    12.68  &    12.61  &     0.08\\ 
    155   & \nodata   &             \nodata   &     0.77  &       \nodata     &    0.77  &     6.85  &     7.17  &    -0.32\\ 
    156   & \nodata   &             \nodata   &     0.86  &       \nodata     &    0.86  &     8.10  &     8.37  &    -0.27\\ 
    162   & \nodata   &             \nodata   &     0.85  &       \nodata     &    0.85  &    10.52  &    10.73  &    -0.21\\ 
    163   & A3   &             \nodata   &       \nodata  &       1.14   &    1.14  &    10.62  &    10.55  &     0.07\\ 
    166   & \nodata   &             \nodata   &     0.99  &       \nodata     &    0.99  &    10.73  &    10.95  &    -0.22\\ 
    176   & \nodata   &             \nodata   &     0.93  &       \nodata     &    0.93  &     8.46  &     8.73  &    -0.27\\ 
    177   & \nodata   &             \nodata   &     0.85  &       \nodata     &    0.85  &     7.37  &     7.68  &    -0.31\\ 
    179   & G0   &             \nodata   &       \nodata  &       0.54   &    0.54  &    12.26  &    11.69  &     0.58\\ 
    180*  & B5   &             1.1  &     \nodata    &       1.13   &    1.13  &    11.18  &    11.29  &    -0.10\\ 
    182   & A4   &             0.6: &     \nodata    &       0.75   &    0.75  &    13.78  &    13.68  &     0.11\\ 
    183*  & B2   &             1.4  &     1.21  &       1.00   &    1.21  &     6.81  &     7.12  &    -0.31\\ 
    184*  & B8   &               1  &     \nodata    &       0.94   &    0.94  &    12.11  &    12.22  &    -0.10\\ 
    185   & \nodata   &             \nodata   &     0.83  &       \nodata     &    0.83  &     7.07  &     7.37  &    -0.30\\ 
    186   & \nodata   &             \nodata   &     0.78  &       \nodata     &    0.78  &     8.82  &     9.11  &    -0.29\\ 
    188   & A0   &             1.5  &     \nodata    &       0.80   &    0.80  &    13.15  &    13.20  &    -0.05\\ 
    189   & \nodata   &              \nodata  &     1.16  &              &    1.16  &    10.10  &    10.35  &    -0.25\\ 
    190   & B8   &             1.0: &     \nodata    &       0.76   &    0.76  &    12.56  &    12.66  &    -0.10\\ 
    195   & B5   &             0.9  &     \nodata    &       0.85   &    0.85  &    11.98  &    12.09  &    -0.10\\ 
    197   & \nodata   &             \nodata   &     0.86  &       \nodata     &    0.86  &     5.87  &     6.19  &    -0.32\\ 
    199   & \nodata   &             \nodata   &     0.78  &       \nodata     &    0.78  &     7.90  &     8.18  &    -0.28\\ 
    201*  & A0   &            1.25  &     \nodata    &       0.75   &    0.75  &    13.63  &    13.64  &     0.00\\ 
    206   & B5   &             1.1  &     0.82  &       0.72   &    0.82  &    10.75  &    10.96  &    -0.20\\ 
    207   & F3   &             \nodata   &       \nodata  &       0.60   &    0.60  &    11.42  &    11.07  &     0.35\\ 
    208   & \nodata   &             \nodata   &     0.98  &       \nodata     &    0.98  &    10.68  &    10.90  &    -0.22\\ 
    209   & B5   &             1.4  &     \nodata    &       0.94   &    0.94  &    12.29  &    12.40  &    -0.10\\ 
    223   & \nodata   &             \nodata   &     1.09  &       \nodata     &    1.09  &     8.86  &     9.12  &    -0.27\\ 
    224   & A9   &             \nodata   &       \nodata  &       0.55   &    0.55  &    13.06  &    12.79  &     0.28\\ 
    228   & B8   &             0.9  &     \nodata    &       1.00   &    1.00  &    11.78  &    11.88  &    -0.10\\ 
    229   & F5   &             \nodata   &       \nodata  &       0.71   &    0.71  &    12.59  &    12.15  &     0.44\\ 
    234   & F6   &             \nodata   &       \nodata  &       0.62   &    0.62  &    12.95  &    12.49  &     0.46\\ 
    237   & F2   &             \nodata   &       \nodata  &       0.65   &    0.65  &    13.13  &    12.79  &     0.35\\ 
    244*  & B5   &             0.9  &     0.68  &       0.55   &    0.68  &     9.20  &     9.44  &    -0.24\\ 
    249   & A0   &            1.75  &     \nodata    &       0.65   &    0.65  &    13.57  &    13.63  &    -0.05\\ 
    251   & A1   &             0.8  &     \nodata    &       0.76   &    0.76  &    13.21  &    13.20  &     0.01\\ 
    255   & B5   &             0.8  &     \nodata    &       0.69   &    0.69  &    12.62  &    12.72  &    -0.10\\ 
    262   & A1   &             0.9: &     \nodata    &       0.77   &    0.77  &    12.85  &    12.84  &     0.01\\ 
    271*  & F8   &             \nodata   &     \nodata    &       1.03   &    1.03  &    12.77  &    12.28  &     0.50\\ 
    272*  & F5   &             \nodata   &     \nodata    &       0.51   &    0.51  &    13.98  &    13.54  &     0.44\\ 
    288   & A1   &             0.9  &     \nodata    &       0.65   &    0.65  &    13.39  &    13.37  &     0.03\\ 
    293   & A7   &             \nodata   &     \nodata    &       0.84   &    0.84  &    12.76  &    12.57  &     0.20\\ 
    297   & A2   &             1.3  &     \nodata    &       0.94   &    0.94  &    12.89  &    12.85  &     0.05\\ 
    306   & F0   &             \nodata   &     \nodata    &       0.52   &    0.52  &    12.22  &    11.92  &     0.30\\ 
    307   & F6   &             \nodata   &     \nodata    &       0.84   &    0.84  &    13.70  &    13.25  &     0.46\\ 
    308   & G5   &             \nodata   &     \nodata    &       0.84   &    0.84  &    13.68  &    13.09  &     0.60\\ 
    314   & B8   &            1.05  &     \nodata    &       1.05   &    1.05  &    11.28  &    11.39  &    -0.10\\ 
    324   & A3   &            \nodata    &     \nodata    &       0.75   &    0.75  &    11.35  &    11.28  &     0.08\\ 
    327   & F0   &            \nodata    &     \nodata    &       0.60   &    0.60  &    13.89  &    13.59  &     0.30\\ 
    332   & B8   &             1.2  &     \nodata    &       0.91   &    0.91  &    11.81  &    11.92  &    -0.10\\ 
    340*  & B5   &             1.3  &     \nodata    &       0.66   &    0.66  &    11.36  &    11.46  &    -0.10\\ 
    343*  & A0   &               1  &     \nodata    &       1.16   &    1.16  &    11.67  &    11.69  &    -0.03\\ 
    349   & G5   &             \nodata   &       \nodata  &       0.78   &    0.78  &    13.81  &    13.21  &     0.60\\ 
    356   & F5   &             \nodata   &       \nodata  &       0.56   &    0.56  &    14.59  &    14.15  &     0.44\\ 
    361   & A6   &             \nodata   &       \nodata  &       0.62   &    0.62  &    13.01  &    12.83  &     0.18\\ 
    362   & A0   &             1.3  &       \nodata  &       0.75   &    0.75  &    12.79  &    12.82  &    -0.02\\ 
    369   & F5   &             \nodata   &       \nodata  &       0.62   &    0.62  &    13.71  &    13.27  &     0.44\\ 
    370   & \nodata   &             \nodata   &     0.89  &       \nodata     &    0.89  &    10.66  &    10.88  &    -0.22\\ 
    374   & A2   &             1.1  &     \nodata    &       0.68   &    0.68  &    13.67  &    13.62  &     0.05\\ 
    375   & A2   &             1.4  &     \nodata    &       0.82   &    0.82  &    12.79  &    12.77  &     0.03\\ 
    377   & B5   &             1.5  &     \nodata    &       0.74   &    0.74  &    12.80  &    12.91  &    -0.10\\ 
    386*  & B5   &             1.4  &     \nodata    &       1.03   &    1.03  &    11.10  &    11.21  &    -0.10\\ 
    398   & B5   &             1.3  &     \nodata    &       0.51   &    0.51  &    13.94  &    14.04  &    -0.10\\ 
    403   & \nodata   &             \nodata   &     0.87  &       \nodata     &    0.87  &    10.49  &    10.65  &    -0.15\\ 
    405   & G5   &             \nodata   &       \nodata  &       0.50   &    0.50  &    14.09  &    13.49  &     0.60\\ 
    406   & B5   &             1.3  &     0.71  &       0.65   &    0.71  &    10.75  &    10.92  &    -0.17\\ 
    414   & F6   &             \nodata   &       \nodata  &       0.59   &    0.59  &    12.72  &    12.26  &     0.46\\ 
    420   & A9   &             2.1::&       \nodata  &       0.88   &    0.88  &    12.04  &    11.77  &     0.27\\ 
    423   & \nodata   &             \nodata   &     0.66  &       \nodata     &    0.66  &     8.62  &     8.86  &    -0.24\\ 
    431   & F3   &             \nodata   &       \nodata  &       0.71   &    0.71  &    12.75  &    12.40  &     0.35\\ 
    437   & A2   &             1.1  &     \nodata    &       0.74   &    0.74  &    13.19  &    13.15  &     0.05\\ 
    440   & B5   &            1.35  &     \nodata    &       0.75   &    0.75  &    11.60  &    11.71  &    -0.10\\ 
    444   & F8   &             \nodata   &       \nodata  &       0.53   &    0.53  &    13.28  &    12.77  &     0.52\\ 
    446   & A0   &             \nodata   &       \nodata  &       0.87   &    0.87  &    11.60  &    11.65  &    -0.05\\ 
    447   & F5   &             \nodata   &       \nodata  &       0.57   &    0.57  &    13.78  &    13.34  &     0.44\\ 
    448   & A0   &             \nodata   &       \nodata  &       0.66   &    0.66  &    13.90  &    13.92  &    -0.02\\ 
    449   & A1   &            0.95  &     \nodata    &       0.69   &    0.69  &    13.16  &    13.15  &     0.01\\ 
    451   & A0   &             1.4  &     \nodata    &       0.72   &    0.72  &    13.04  &    13.05  &     0.00\\ 
    476   & A3   &             1.1  &     \nodata    &       0.73   &    0.73  &    12.65  &    12.58  &     0.08\\ 
    484   & A2   &             1.9  &     \nodata    &       0.83   &    0.83  &    13.16  &    13.12  &     0.05\\ 
    489   & B5   &               1  &     0.91  &       0.83   &    0.91  &    10.92  &    11.11  &    -0.19\\ 
    496   & A8   &             \nodata   &       \nodata  &       0.61   &    0.61  &    12.81  &    12.54  &     0.27\\ 
    504   & B6   &             1.1  &     0.81  &       0.71   &    0.81  &    11.23  &    11.44  &    -0.21\\ 
    509   & F7   &             \nodata   &       \nodata  &       0.65   &    0.65  &    13.38  &    12.91  &     0.48\\ 
    514   & B5   &               1  &     0.99  &       0.87   &    0.99  &    10.60  &    10.82  &    -0.22\\ 
    517   & \nodata   &             \nodata   &    1.06   &       \nodata     &    1.06  &     9.26  &     9.45  &    -0.19\\ 
    519   & G5   &             \nodata   &       \nodata  &       0.53   &    0.53  &    13.29  &    12.70  &     0.60\\ 
    520*  & A1   &             1.2  &     \nodata    &       0.96   &    0.96  &    11.90  &    11.89  &     0.01\\ 
    526   & A1   &             1.5  &     \nodata    &       1.01   &    1.01  &    11.63  &    11.63  &     0.01\\ 
    527   & A4   &             \nodata   &       \nodata  &       0.94   &    0.94  &    11.92  &    11.80  &     0.13\\ 
    538   & A0   &             1.4  &     \nodata    &       0.95   &    0.95  &    12.05  &    12.08  &    -0.02\\ 
    551   & \nodata   &             \nodata   &     1.30  &       \nodata     &    1.30  &     9.17  &     9.41  &    -0.24\\ 
    555   & G5   &             \nodata   &       \nodata  &       0.55   &    0.55  &    12.73  &    12.39  &     0.35\\ 
    556   & B5   &             1.1  &     \nodata    &       0.94   &    0.94  &    11.26  &    11.37  &    -0.10\\ 
    559   & B5   &             1.3  &     \nodata    &       0.94   &    0.94  &    11.07  &    11.18  &    -0.10\\ 
    562   & F5   &             \nodata   &       \nodata  &       0.70   &    0.70  &    12.90  &    12.46  &     0.44\\ 
    565   & B8   &            1.65  &     \nodata    &       1.14   &    1.14  &    11.00  &    11.11  &    -0.11\\ 
    566   & F2   &             \nodata   &     \nodata    &       0.79   &    0.79  &    11.79  &    11.44  &     0.35\\ 
    577   & A0   &             1.1  &     \nodata    &       0.87   &    0.87  &    12.60  &    12.62  &    -0.02\\ 
    579   & A0   &             1.6  &     \nodata    &       0.94   &    0.94  &    12.03  &    12.06  &    -0.02\\ 
    582   & A9   &             \nodata   &       \nodata  &       0.89   &    0.89  &    11.79  &    11.51  &     0.28\\ 
    590   & A0   &            1.15  &     \nodata    &       0.69   &    0.69  &    13.19  &    13.21  &    -0.02\\ 
    594   & A0   &             1.3  &     \nodata    &       0.89   &    0.89  &    12.04  &    12.06  &    -0.02\\ 
    595   & G5   &             \nodata   &       \nodata  &       0.62   &    0.62  &    13.93  &    13.33  &     0.60\\ 
    597   & A5   &             1.4  &     \nodata    &       0.65   &    0.65  &    12.67  &    12.53  &     0.15\\ 
    601   & B5   &             1.7  &     \nodata    &       0.72   &    0.72  &    12.57  &    12.68  &    -0.10\\ 
    605   & \nodata   &             \nodata   &     0.87  &       \nodata     &    0.87  &     9.44  &     9.69  &    -0.25\\ 
    642*  & A8   &             \nodata   &     \nodata    &       0.61   &    0.61  &    12.75  &    12.48  &     0.27\\ 
    652*  & B5   &             1.3  &     \nodata    &       0.94   &    0.94  &    11.49  &    11.60  &    -0.10\\ 
    655   & A4   &             \nodata   &       \nodata  &       0.68   &    0.68  &    12.65  &    12.54  &     0.11\\ 
    680   & G5   &             \nodata   &       \nodata  &       0.86   &    0.86  &    13.79  &    13.19  &     0.60\\ 
    681   & B5   &            1.25  &     \nodata    &       0.67   &    0.67  &    12.12  &    12.22  &    -0.10\\ 
    682*  & A3   &             \nodata   &       \nodata  &       0.81   &    0.81  &    12.20  &    12.12  &     0.08\\ 
    690   & K0   &             \nodata   &       \nodata  &       0.55   &    0.55  &    12.97  &    12.28  &     0.70\\ 
    718   & \nodata   &             \nodata   &     0.82  &       \nodata     &    0.82  &     9.56  &     9.72  &    -0.16\\ 
    719   & A0   &            1.25  &     \nodata    &       0.82   &    0.82  &    12.90  &    12.93  &    -0.02\\ 
    734   & \nodata   &             \nodata   &     0.92  &       \nodata     &    0.92  &     9.38  &     9.53  &    -0.15\\ 
    740*  & A4   &             1.6  &     \nodata    &       0.82   &    0.82  &    12.93  &    12.81  &     0.13\\ 
    750   & A0   &             1.3  &     \nodata    &       0.92   &    0.92  &    12.08  &    12.11  &    -0.02\\ 
    754   & A0   &             1.2  &     \nodata    &       0.79   &    0.79  &    11.03  &    11.05  &    -0.02\\ 
    759   & B5   &             1.4  &     \nodata    &       1.12   &    1.12  &    11.08  &    11.19  &    -0.10\\ 
    824   & \nodata   &             \nodata   &     0.81  &       \nodata     &    0.81  &    10.25  &    10.48  &    -0.23\\ 
    826*  & A3   &             \nodata   &       \nodata  &       0.52   &    0.52  &    14.31  &    14.20  &     0.11\\ 
    831   & \nodata   &             \nodata   &     1.04  &       \nodata     &    1.04  &    10.58  &    10.76  &    -0.18\\ 
    854   & F6   &             \nodata   &       \nodata  &       0.69   &    0.69  &    12.84  &    12.38  &     0.46\\ 
    859   & \nodata   &             \nodata   &     1.09  &       \nodata     &    1.09  &    10.08  &    10.28  &    -0.20\\ 
    871   & G5   &             \nodata   &       \nodata  &       0.63   &    0.63  &    14.64  &    14.05  &     0.60\\ 
    872   & \nodata   &             \nodata   &     0.76  &       \nodata     &    0.76  &     8.47  &     8.73  &    -0.27\\ 
    876   & B5   &               1  &     \nodata    &       0.69   &    0.69  &    12.06  &    12.16  &    -0.10\\ 
    877   & B5   &             1.6  &     \nodata    &       1.06   &    1.06  &    11.19  &    11.29  &    -0.10\\ 
    879   & G5   &             \nodata   &       \nodata  &       0.54   &    0.54  &    13.20  &    12.61  &     0.60\\ 
    884   & \nodata   &             \nodata   &     0.91  &       \nodata     &    0.91  &    10.59  &    10.79  &    -0.20\\ 
    887   & \nodata   &             \nodata   &     0.89  &       \nodata     &    0.89  &     9.89  &    10.12  &    -0.23\\ 
    888   & \nodata   &             \nodata   &     0.75  &       \nodata     &    0.75  &    10.36  &    10.57  &    -0.21\\ 
    890   & A2   &             0.7: &       \nodata  &       0.56   &    0.56  &    14.16  &    14.11  &     0.05\\ 
    891   & B5   &            1.3   &     1.09  &       1.03   &    1.09  &    10.38  &    10.54  &    -0.17\\ 
    896   & \nodata   &           1.3    &     0.87  &       \nodata     &    0.87  &    11.08  &    11.26  &    -0.18\\ 
    898   & A1   &           0.8:   &     \nodata    &       0.65   &    0.65  &    11.96  &    11.96  &     0.01\\ 
    912*  & B5   &               1  &     \nodata    &       1.10   &    1.10  &    12.25  &    12.35  &    -0.10\\ 
    913   & A0   &             0.8  &     \nodata    &       0.58   &    0.58  &    14.34  &    14.36  &    -0.02\\ 
    919   & A2   &             1.3  &     \nodata    &       0.78   &    0.78  &    13.69  &    13.66  &     0.03\\ 
    921   & A5   &             \nodata   &      \nodata   &       0.79   &    0.79  &    13.50  &    13.35  &     0.15\\ 
    924   & \nodata   &             \nodata   &     0.86  &       \nodata     &    0.86  &     9.67  &     9.91  &    -0.24\\ 
    927   & F2   &             \nodata   &       \nodata  &       0.58   &    0.58  &    12.69  &    12.34  &     0.35\\ 
    928   & A0   &             1.2  &     \nodata    &       0.84   &    0.84  &    11.34  &    11.40  &    -0.05\\ 
    929   & F2   &               1::&     \nodata    &       0.77   &    0.77  &    13.62  &    13.27  &     0.35\\ 
    930   & \nodata   &             \nodata   &    1.05   &       \nodata     &    1.05  &     9.13  &     9.35  &    -0.21\\ 
    937   & A5   &             \nodata   &       \nodata  &       0.85   &    0.85  &    13.18  &    13.04  &     0.15\\ 
    939   & \nodata   &             \nodata   &     1.02  &       \nodata     &    1.02  &     9.45  &     9.68  &    -0.23\\ 
    947   & \nodata   &             \nodata   &     1.19  &       \nodata     &    1.19  &     9.75  &    10.00  &    -0.26\\ 
    948   & F5   &             \nodata   &       \nodata  &       0.66   &    0.66  &    13.47  &    13.03  &     0.44\\ 
    966   & A2   &             \nodata   &       \nodata  &       0.76   &    0.76  &    10.79  &    10.74  &     0.05\\ 
    970   & A8   &             \nodata   &       \nodata  &       0.62   &    0.62  &    12.76  &    12.49  &     0.27\\ 
    980   & \nodata   &             \nodata   &     1.30  &       \nodata     &    1.30  &     6.96  &     7.25  &    -0.29\\ 
    984   & \nodata   &             \nodata   &     1.01  &       \nodata     &    1.01  &     9.33  &     9.56  &    -0.24\\ 
    987   & B7   &            1.55  &     \nodata    &       1.00   &    1.00  &    11.92  &    12.02  &    -0.10\\ 
    991   & B8   &             1.2  &     \nodata    &       0.88   &    0.88  &    11.33  &    11.43  &    -0.10\\ 
\enddata
\tablenotetext{a}{Table will also be available electronically at the
Journal.}
\tablenotetext{b}{An additional asterisk (*) denotes that we determine
that the object has an IR excess.}
\end{deluxetable}
\clearpage

\begin{deluxetable}{lp{11cm}}
\tablecaption{Stellar Properties Inconsistent with Membership \label{tab:rejects}}
\tablewidth{0pt}
\tablehead{
\colhead{No.\ of stars} & \colhead{Reason for Rejection as Members} }
\startdata
40 & Reddening derived from spectral classification inconsistent with
membership\\
27 & $B-V > 1.9$; even if we apply the maximum reddening correction, 
these stars are redder than the reddest likely members plotted in Figure 2 and
with this maximum correction applied, they also fall above the track for 2 \msun\
stars, which marks the upper boundary for IC 1805 members.\\
51 & $B-V > 1.5$ and $V<15$; for any assumed value of the reddening 
these stars populate a region of the HR diagram inconsistent 
with the age of IC 1805.  These stars are probably field giants 
(e.g., Hillenbrand \etal\ 1993)\\
16 & Positions in the $J$, $J-H$ diagram, after correction for the mean 
reddening of IC 1805, are inconsistent with membership\\
12 & $B-V$ too blue after correction for mean reddening\\
5 & Reddening derived by $Q$-method $>$ 1.3\\
\enddata
\end{deluxetable}

In summary, there are:  1) 229 members based on reddening values
consistent with  those observed for early-type members of the cluster;
2) 151 stars that are unlikely to be  members based on either
reddening or position in the HR diagram; and 3) 31 stars  likely to be
members because they have IR excesses.  Of the 802 stars observed in
the  IRAC bands, we have no membership information for 391 of them.  A
method for  estimating what fraction of these are actually members is
discussed in \S\ref{sec:statmembers}.

\subsection{Masses and Ages of Sample}
\label{sec:massage}

\begin{figure*}[tbp]
\epsscale{0.75}
\plotone{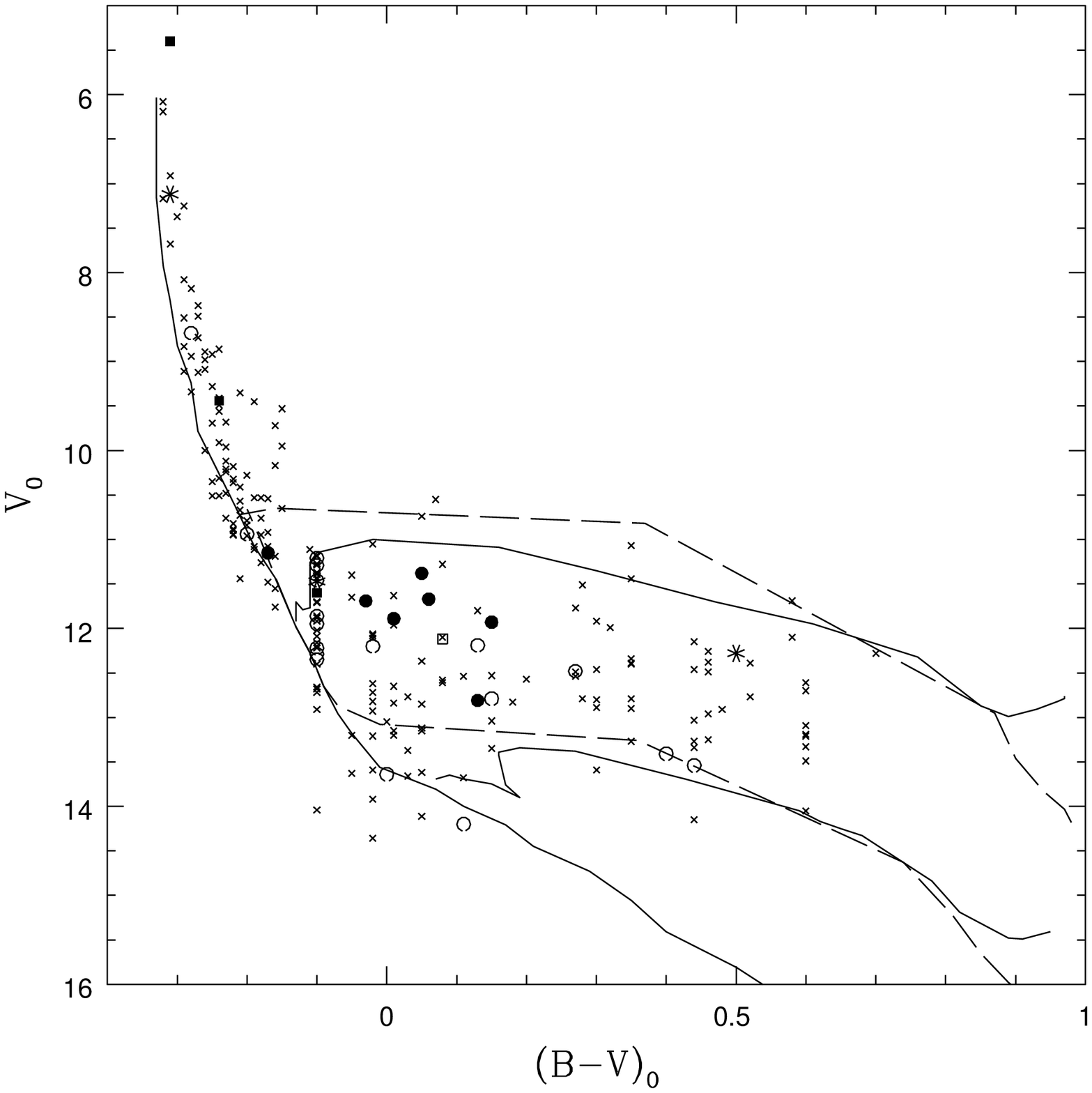}
\caption{CMD for likely 
members of IC 1805.  Solid lines  indicated the locations of
the ZAMS and pre-main sequence evolutionary tracks for stars  of 2 and
4 \msun.  The dashed lines are isochrones for 500,000 and 5 Myr. 
Filled circles  represent stars classified as AeBe stars; open circles
are stars with optically thin  emission only in the IRAC bands, and
usually only at [8]; asterisks represent stars with  optically thin
emission at all observed wavelengths; filled squares are stars with
excess  emission at [24] but not at shorter wavelengths; and open
squares are stars with  optically thin emission at $JHK_s$ but
optically thick emission either in the IRAC bands or at  [24]. 
Crosses are stars with no IR excess.}
\label{fig:cmd}
\end{figure*}

Our data provide the basis for assessing the range of disk
evolutionary states for  intermediate and high mass stars in IC 1805. 
The first step is to estimate the masses of  the 229 likely cluster
members selected according to the above criteria. We placed  these
stars in an HR diagram by correcting their colors for the reddening
listed in the  sixth column of Table~\ref{tab:members}. The adopted
reddening law is given in Table~\ref{tab:reddening}, normalized to an 
absorption, $A_K$, of 1.0 mag at 2.2 \mum.  This law is an average of
the  determinations by Rieke \& Lebofsky (1985), Indebetouw \etal\
(2005), and Flaherty (2007).  At  IR wavelengths, the biggest
difference in these various determinations is 0.1 mag  at the 4.5
\mum\ Spitzer band.  Also listed in Table~\ref{tab:reddening} is the
average reddening in IC 1805 estimated by  Massey \etal\ (1995a) of
$E(B-V) = 0.87$, who found $A_{V} = 3.1 \times E(B-V)$.  We have used
our adopted reddening law to derive the average reddening for the IR
wavelengths, which  we apply to those stars for which spectra, and
thus individual reddening estimates, are  unavailable (see below).

The color-magnitude diagram for the members of IC 1805 for which we
have  measurements of the individual reddening is shown in
Figure~\ref{fig:cmd}.  Thirty-two of these 229  members have infrared
excesses (see \S\ref{sec:irx}) and are indicated in the Figure.  Also 
shown are evolutionary tracks, isochrones, and the zero-age main
sequence (ZAMS) from Siess, Dufour, \&  Forestini (2000) up to 7
\msun\ and from Schaller \etal\ (1992) for more massive stars.   For
the masses included in the tabulation by Siess \etal, we used their
conversions to  colors and magnitudes.  For the more massive stars, we
used the conversion from  temperature to $B-V$ from Allen's
Astrophysical Quantities and the bolometric corrections  from Massey
\etal\ (2005).  As the Figure indicates, the pre-main sequence stars
in our  sample have masses in the range 2-4 \msun. Stars more massive
than 4 \msun\ have  already reached the main sequence.

\begin{deluxetable}{lll}
\tablecaption{Adopted Reddening Law\label{tab:reddening}}
\tablewidth{0pt}
\tablehead{
\colhead{Filter} & \colhead{Reddening Law} & \colhead{IC 1805 Average
Reddening} }
\startdata
$U$ (0.36 \mum) & 13.67 & 4.13\\
$B$ (0.44 \mum) & 11.82 & 3.57\\
$V$ (0.55 \mum) & 8.93  & 2.70\\
$J$ (1.24 \mum) & 2.45  & 0.74\\
$H$ (1.65 \mum) & 1.55  & 0.47\\
$K_s$ (2.17 \mum) &  1.00 & 0.3 \\
 3.6 \mum\ & 0.6 & 0.18  \\
 4.5 \mum\ & 0.49 & 0.15 \\
 5.8 \mum\ & 0.46 & 0.14 \\
 8 \mum\ & 0.46 & 0.14 \\
 24 \mum\ & 0.48 & 0.14 \\
\enddata

\end{deluxetable}

Massey \etal\ (1995a) estimated an age range of 1-3 Myr for the
massive stars based  on the fact that these stars still lie very close
to the ZAMS.  The isochrones plotted in  Figure~\ref{fig:cmd} suggest
a somewhat larger age range of 0.5-5 Myr for the pre-main sequence 
2-4 \msun\ stars, with the 5 Myr limit set by the limiting magnitude
of the survey.  The use  of position in the HR diagram to determine
ages for pre-main-sequence intermediate  mass stars is, however,
highly problematic.  For low mass stars, ages are estimated  from a
zero point called the ``birthline.''  The birthline is essentially the
mass-radius  relationship for pre-main sequence stars that have
completed the main accretion phase  and have begun their quasi-static
contraction toward the main sequence.  For stars in  the 2-4 \msun\
range studied here, the birthline is critically dependent on the
accretion  rate during the infall phase.  Because accretion rates
during this phase may vary widely, an ensemble of young stars in this
mass range may initiate  their quasi-static contraction from a variety
of initial values of mass and radius rather than from a single
well-defined birthline.

Empirically, the problem of determining ages for stars in this mass
range is illustrated by the fact that the apparent ages of stars as
determined from positions in the HR diagram  appear to increase with
increasing mass.  For example, in the Orion cluster, the age of  the 2
\msun\ stars is about 5 times older than the age of 0.5 \msun\ stars. 
Depending on  the calibration chosen, the 2 \msun\ stars appear to
have an age of about 5 Myr old  rather than the 1 Myr age estimated for
the lower mass stars (Hillenbrand 1997).  A similar  effect is seen in
W5, which is a region that closely resembles IC 1805 (Koenig \& Allen 
2010).

Given these uncertainties, we will assume that the age of our sample
is similar to that of  the massive stars -- 1-3 Myr.

\subsection{Cluster Properties}

\begin{figure*}[tbp]
\epsscale{0.9}
\plotone{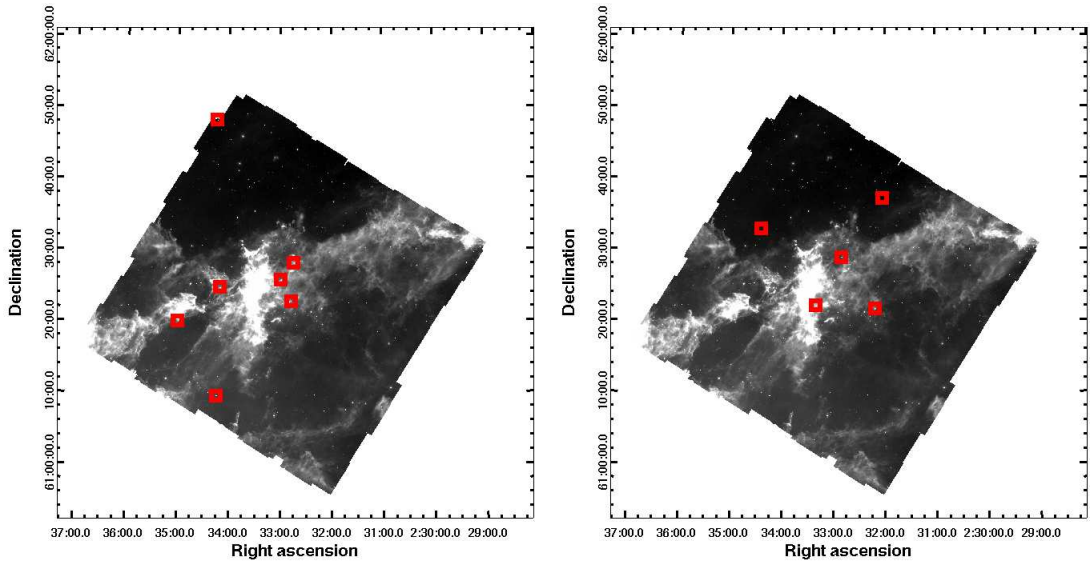}
\caption{Locations of various SED categories (see \S\ref{sec:SED})
overlaid on the 8 \mum\ image, part 1.  Left: optically thick excess
emission from the near-infrared through [24] (i.e., Herbig Ae/Be
stars); right: optically thin excesses in all of the observed IR
bands. }
\label{fig:backgroundism1}
\end{figure*}
\begin{figure*}[tbp]
\epsscale{0.9}
\plotone{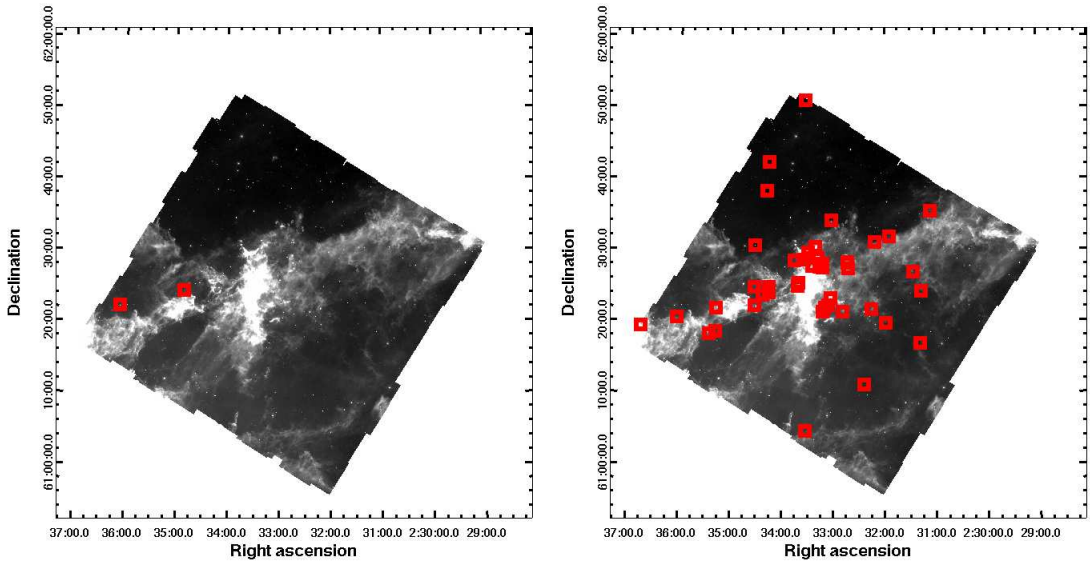}
\caption{Locations of various SED categories (see \S\ref{sec:SED})
overlaid on the 8 \mum\ image, part 2.  Left: optically thin emission
in the near-IR bands and optically thick emission at 24 \mum\ (i.e.,
thin/thick); right: no excess emission in the near-IR but optically
thin excess emission at [8] and occasionally at [5.8] and once at [24]
(i.e., empty/thin).  }
\label{fig:backgroundism2}
\end{figure*}
\begin{figure*}[tbp]
\epsscale{0.9}
\plotone{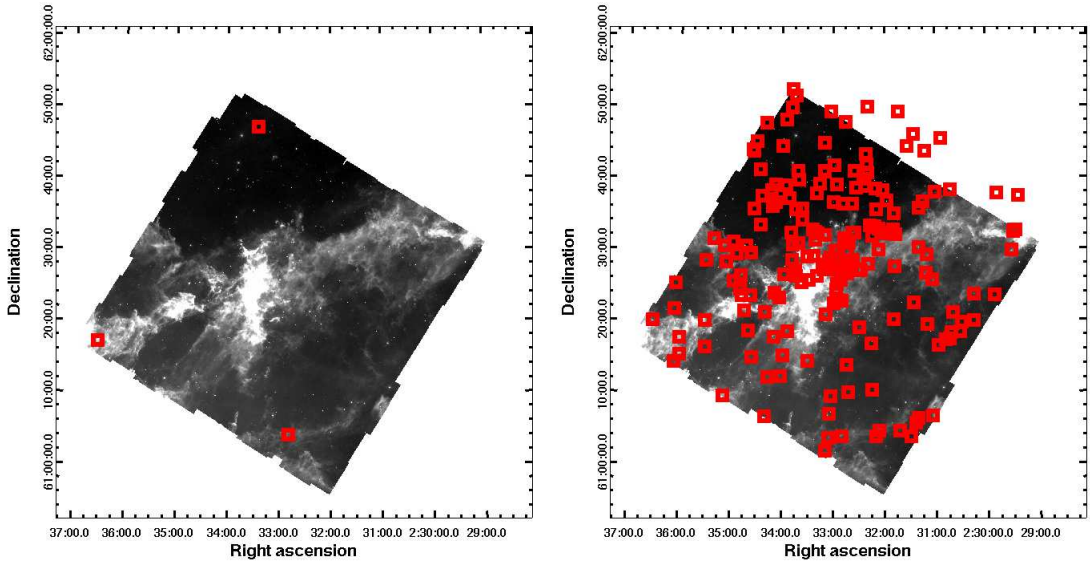}
\caption{Locations of various SED categories (see \S\ref{sec:SED})
overlaid on the 8 \mum\ image, part 3.  Left: optically thick emission
detected only at [24] (i.e., empty/thick); right: no excess.  }
\label{fig:backgroundism3}
\end{figure*}

\begin{figure*}[tbp]
\epsscale{0.75}
\plotone{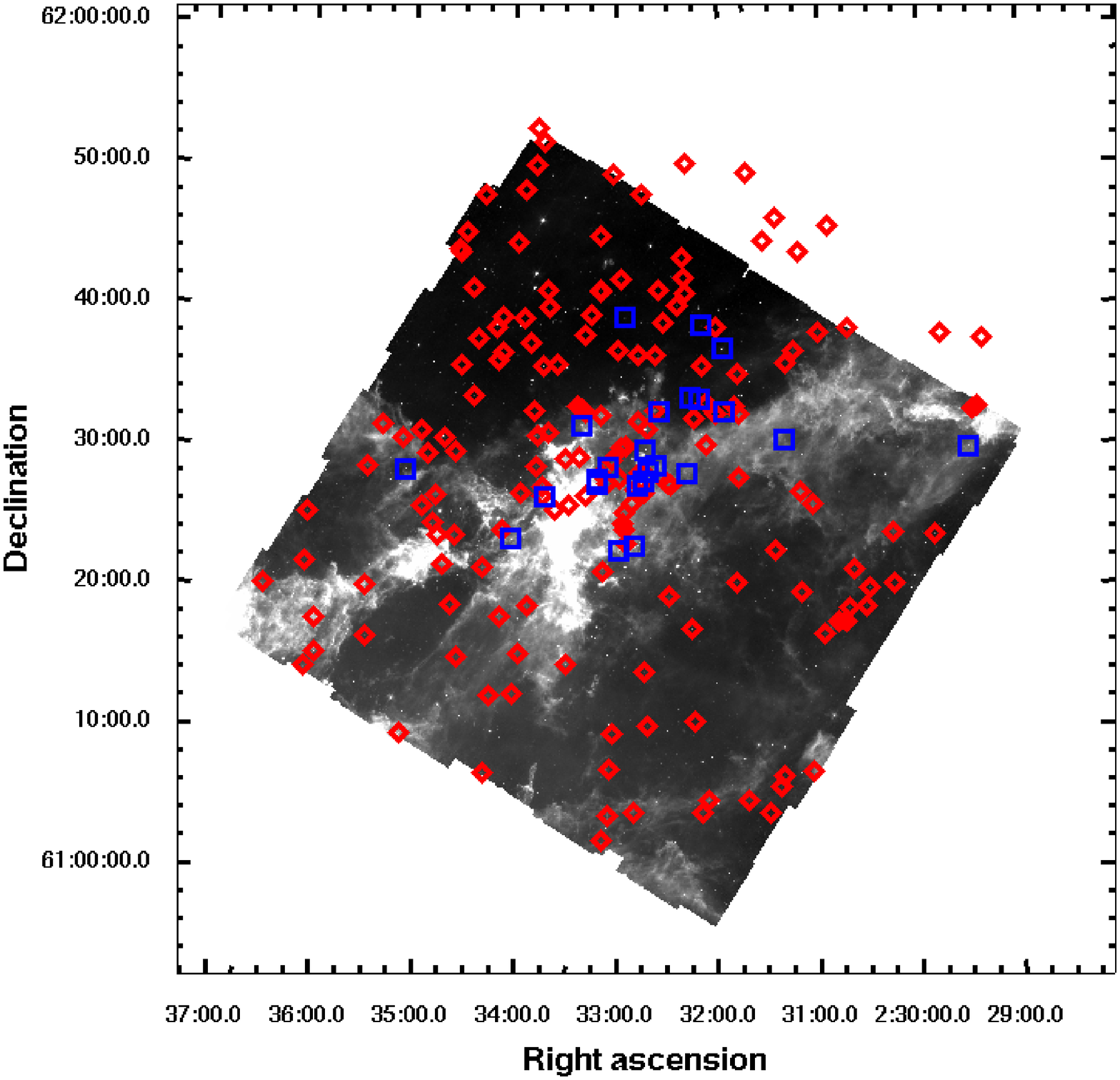}
\caption{Locations of IC 1805 member stars without IR excesses
overlaid on an 8 \mum\ image. The high mass ($M >12$
\msun) stars are blue squares, and the lower mass stars ($M<12$ \msun)
are red diamonds.  Some symbols appear off the edge of the 8 \mum\
map  because measurements exist at other Spitzer bands (e.g., 3.6 and
5.8 \mum).  North is up. The higher mass stars are more concentrated
than the lower mass stars. }
\label{fig:massseg}
\end{figure*}

Figures~\ref{fig:backgroundism1}-\ref{fig:backgroundism3} show the
locations projected on the sky of the IC 1805 stars with IR excesses,
broken down in the categories discussed in \S\ref{sec:SED} below.  
Most of these stars lie along the region of bright 8-\mum\
interstellar dust emission  stretching from southeast to northwest
across the image, with a concentration of stars in  the central
region.  There are also, however, stars with IR excesses that fall in
regions  where little dust emission is seen.  

Figure~\ref{fig:massseg} shows a similar plot for the members of IC
1805 listed in Table~\ref{tab:members} and that do not have IR
excesses.  This plot shows that the massive stars ($M>$12 \msun) are
more strongly concentrated than the lower mass stars. Indeed
most of the massive stars are  found in a region that is about
0.2$\arcdeg$ across or, at the 2350 pc distance of IC 1805  (Massey
\etal\ 1995a), $\sim$4 pc in radius. The velocity dispersion of IC
1805 is not  known, but if we adopt 5 km s$^{-1}$ and a typical age of
2 Myr, then massive stars can move  only $\sim$10 pc in that time. 
The whole region covered by this survey where recent star  formation
has taken place is about 0.8$\arcdeg$ across or 16 pc in radius. 
Since the massive  stars could not have traversed this entire distance
in their estimated lifetimes, it seems  likely that the observed mass
segregation was imposed at the time of their formation.   The same
conclusion was reached many years ago by Sagar \etal\ (1988) based on 
proper motion measurements of the stars in IC 1805.  A recent study of
another young  region, W51, similarly finds evidence for a stronger
concentration of YSOs with  $M>8$ \msun\ when compared with those with
$M<5$ \msun\ (Kang \etal\ 2009).

\section{Infrared Excesses}
\label{sec:irx}

\subsection{Selection of Stars with Infrared Excesses}
\label{sec:finalirx}

In order to identify stars with infrared excesses, we assumed that there
is no excess  emission above the photosphere at $J$.  We corrected the
observations at each of the IR  wavelengths for reddening (see
Table~\ref{tab:irx}) and then subtracted the magnitude relative to $J$
appropriate  for the stellar photospheric temperature at the
wavelengths longer than $J$ (see Table~\ref{tab:irx-photosphere}). 
For 32 of the  stars that we found to have excesses, we have estimates
of the reddening from either  the $Q$ method or spectral typing. 
There are an additional 31 stars with IR excesses but  with
indeterminate reddening because they are too cool for the $Q$ method
to be applicable  and no spectra are available.  We have assumed that
these stars are members of IC 1805  by virtue of their IR excesses and
for them have adopted the mean reddening of IC  1805 (i.e., A$_V =
$2.7). Fortunately, as Table~\ref{tab:reddening} shows, the reddening
in the infrared in IC 1805 is small,  and the reddening is nearly
independent of wavelength for the Spitzer bands.  For the  range of
reddening observed for members of IC 1805 ($0.5 < E(B-V) < 1.30$),
adoption of  the mean reddening leads to a maximum error of 0.3 mag in
the color $J-[24]$ and $<$ 0.1 mag for [3.6]$-$[24].

For photospheric flux densities, we adopted the black body models calculated
by Koenig \&  Allen (2010) and the temperatures derived from the
visible colors after applying  individual reddening values for the
stars for which measurements are available.  For the  stars for which
we do not have a measurement of the reddening, we have applied the 
mean reddening values and assumed a temperature of \teff = 9000.  The
likely range in  temperatures of the stars with unknown reddening is
$\sim$14000 K (the $Q$ method is valid  for all of the hotter stars)
to $\sim$6000 K (the latest types observed are around G5).  Given 
this range, the error in $J-[8]$ as a result of the difference between
the actual stellar  temperature and 9000 K is at most about 0.3 mag. 
The error in $J-[24]$ as a result of  uncertainty in the temperature
reaches a maximum of 0.4 mag for the coolest stars in  our sample.  

Given the combination of uncertainties in the measured flux densities, the
reddening, and the  intrinsic flux density of the photosphere, we have adopted
the conservative criteria that the  observed IR excess must be $>$0.5
mag at 8 \mum\ or 1.0 mag at 24 \mum\ in order for a star to be
classified as having an IR excess.

The identification of stars with IR excesses is fairly insensitive to
the corrections applied  for reddening and for the photospheric flux
density. In a preliminary reconnaissance of the data, we used plots of
\ks\ vs.\ \ks$-$[IRAC band] to identify stars that lay outside the the
scatter attributable to photometric errors.  The same stars were
identified from both procedures, with the exception that the more
conservative criterion adopted here, namely  that the IR excess had to
exceed 0.5 mag after correction, eliminated 6 stars (Nos.\ 1,  62, 99,
250, 888, and 891).

Table~\ref{tab:irx} lists the extinction-corrected colors in
magnitudes relative to the $J$ band, after correction for 
interstellar reddening.  Table~\ref{tab:irx-photosphere} gives the
magnitudes relative to $J$ after correction for  both interstellar
reddening and photospheric flux density. For the stars without an 
individual estimate of reddening, we applied the mean reddening given in
Table~\ref{tab:reddening} and assumed a temperature of 9000 K, as
described; temperatures are in Table~\ref{tab:irx-photosphere}.
For the other stars, those for which we have
individual  measurements of the reddening from spectroscopy or the 
$Q$ method, we can obtain an estimate of
the stellar temperature, and those temperatures are listed in
Table~\ref{tab:irx-photosphere}.

\begin{deluxetable}{llllllllll}
\tablecaption{ Extinction-Corrected Colors of IC 1805 Members with Infrared Excesses: Infrared
Excesses Relative to $J$ \label{tab:irx}}
\tabletypesize{\tiny}
\rotate
\tablewidth{0pt}
\tablehead{\colhead{Star No.} & \colhead{Excess Type} &
\colhead{$J_0$} & \colhead{($J-H$)$_0$} & \colhead{($J-K_s$)$_0$} &
\colhead{($J-$[3.6])$_0$} & \colhead{($J-$[4.5])$_0$} &
\colhead{($J-$[5.8])$_0$} & \colhead{($J-$[8])$_0$} &
\colhead{($J-$[24])$_0$} \\  \colhead{} & \colhead{} & \colhead{(mag)} &
\colhead{(mag)} & \colhead{(mag)} & \colhead{(mag)} & \colhead{(mag)}
& \colhead{(mag)} & \colhead{(mag)} & \colhead{(mag)} } 
\startdata
4 & Thin & 11.37 & -0.05 & -0.04 & 0.19 & 0.31 & 0.43 & 0.69 & 1.57 \\
6 & Empty/Thin [8] & 11.76 & 0.03 & 0.04 & 0.04 & 0.05 & 0.32 & 0.81 & $<$5.88 \\
13 & Empty/Thin [8] & 9.02 & -0.15 & -0.25 & -0.29 & -0.26 & -0.15 & 0.77 & $<$7.87 \\
14 & Empty/Thin [8] & 11.94 & 0.24 & 0.21 & 0.17 & 0.14 & 0.26 & 0.88 & $<$6.07 \\
15 & Empty/Thin [8] & 12.14 & -0.14 & -0.11 & -0.07 & 0.02 & 0.07 & 0.78 & $<$9.30 \\
18 & Thick (AeBe) & 10.69 & 0.76 & 1.48 & 2.74 & 3.21 & 3.77 & 4.68 & 8.37 \\
51 & Empty/Thin [8] & 12.23 & 0.18 & 0.13 & 0.09 & 0.04 & 0.18 & 0.88 & $<$5.80 \\
61 & Empty/Thin [8] & 12.63 & 0.11 & 0.00 & -0.05 & -0.04 & 0.29 & 1.01 & $<$7.50 \\
65 & Empty/Thin [5.8] & 12.31 & 0.07 & 0.12 & 0.32 & 0.45 & 0.82 & 1.79 & $<$6.69 \\
66 & Empty/Thin [5.8] & 12.45 & 0.11 & 0.11 & 0.20 & 0.18 & 1.00 & 2.51 & $<$8.61 \\
73 & Empty/Thin [8] & 12.34 & 0.20 & 0.28 & 0.31 & 0.24 & 0.62 & 1.44 & $<$7.87 \\
80 & Empty/Thin [8] & 12.46 & 0.19 & 0.16 & 0.18 & 0.17 & 0.28 & 1.17 & $<$5.10 \\
85 & Empty/Thin [8] & 12.35 & -0.05 & -0.05 & -0.02 & -0.04 & 0.19 & 0.92 & $<$6.52 \\
86 & Empty/Thin [8] & 12.32 & 0.04 & 0.03 & 0.00 & 0.00 & 0.03 & 0.70 & $<$4.95 \\
93 & Empty/Thin [8] & 11.92 & 0.06 & 0.09 & 0.08 & 0.05 & 0.33 & 1.06 & $<$5.40 \\
98 & Empty/Thin [8] & 12.21 & 0.06 & -0.01 & 0.03 & 0.05 & 0.35 & 1.04 & $<$7.08 \\
103 & Empty/Thin [8] & 12.64 & -0.01 & -0.06 & 0.01 & 0.01 & 0.37 & 1.41 & $<$7.51 \\
107 & Thick (AeBe) & 10.95 & 0.30 & 0.83 & 2.05 & 2.57 & 2.94 & 3.49 & 5.19 \\
122 & Thin & 11.43 & 0.47 & 0.82 & 1.61 & 1.88 & 2.17 & 3.22 & 3.99 \\
128 & Thick (AeBe) & 11.27 & 0.30 & 0.68 & 1.69 & 2.18 & 2.60 & 3.35 & 6.19 \\
137 & Thick (AeBe) & 11.49 & 0.26 & 0.74 & 2.07 & 2.74 & 3.53 & 5.00 & 7.58 \\
139 & Empty/Thin [5.8] & 11.31 & 0.11 & 0.05 & 0.14 & 0.22 & 0.66 & 2.03 & $<$7.27 \\
143 & Empty/Thin [8] & 12.19 & 0.22 & 0.29 & 0.40 & 0.53 & 0.64 & 1.45 & $<$7.73 \\
144 & Empty/Thin [24] & 6.04 & -0.13 & -0.20 & \nodata & -0.28 & -0.28 & -0.18 & 1.15 \\
180 & Empty/Thin [8] & 11.52 & -0.05 & -0.11 & -0.13 & -0.12 & 0.08 & 0.87 & $<$5.55 \\
183 & Thin & 7.52 & 0.05 & 0.22 & 0.56 & \nodata & 1.01 & 1.40 & 2.62 \\
184 & Empty/Thin [8] & 12.37 & 0.11 & 0.10 & 0.12 & 0.12 & 0.27 & 0.88 & $<$5.71 \\
201 & Empty/Thin [8] & 13.13 & 0.21 & 0.22 & 0.34 & 0.31 & 0.57 & 0.97 & $<$4.99 \\
244 & Empty/Thick [24] & 9.90 & -0.18 & -0.21 & -0.28 & -0.30 & -0.29 & -0.27 & 3.42 \\
259 & Empty/Thin [8] & 12.94 & 0.13 & 0.14 & 0.12 & 0.07 & 0.25 & 0.77 & $<$6.66 \\
261 & Empty/Thin [8] & 12.98 & 0.12 & 0.07 & 0.26 & 0.26 & 0.65 & 2.01 & $<$5.64 \\
270 & Empty/Thin [8] & 13.16 & 0.13 & 0.08 & 0.10 & 0.07 & 0.47 & 1.42 & $<$5.20 \\
271 & Thin & 11.82 & 0.30 & 0.61 & 1.18 & 1.52 & 1.86 & 2.36 & $<$0.57 \\
272 & Empty/Thin [8] & 12.63 & 0.13 & 0.18 & 0.22 & 0.18 & 0.46 & 1.07 & $<$7.88 \\
275 & Empty/Thin [8] & 12.42 & 0.06 & 0.03 & 0.05 & 0.09 & 0.29 & 0.88 & $<$6.94 \\
279 & Empty/Thin [8] & 12.77 & 0.08 & 0.07 & 0.08 & 0.11 & 0.37 & 1.18 & $<$4.56 \\
294 & Empty/Thin [8] & 13.10 & 0.03 & -0.04 & -0.05 & -0.10 & 0.11 & 1.05 & $<$5.13 \\
334 & Empty/Thin [8] & 14.44 & -0.04 & 0.02 & 0.07 & 0.11 & 0.30 & 0.80 & $<$4.36 \\
340 & Empty/Thin [8] & 11.53 & 0.00 & 0.01 & 0.09 & 0.14 & 0.20 & 0.58 & 3.30 \\
343 & Thick (AeBe) & 12.46 & 0.38 & 1.01 & \nodata & 2.50 & \nodata & 4.57 & 8.17 \\
386 & Empty/Thin [8] & 11.69 & -0.12 & -0.12 & -0.20 & -0.22 & -0.18 & 1.56 & $<$0.17 \\
391 & Empty/Thick [24] & 12.80 & -0.10 & -0.12 & -0.14 & -0.17 & -0.15 & 0.04 & 5.87 \\
461 & Thin & 12.59 & 0.18 & 0.29 & 0.69 & 1.02 & 1.26 & 1.82 & \nodata \\
520 & Thick (AeBe) & 11.49 & 0.20 & 0.61 & \nodata & 2.53 & \nodata & 3.61 & 5.39 \\
536 & Empty/Thin [8] & 12.14 & 0.20 & 0.17 & \nodata & 0.13 & \nodata & 0.81 & $<$9.26 \\
563 & Empty/Thick [24] & 11.98 & -0.04 & -0.08 & \nodata & -0.04 & \nodata & 0.50 & 4.23 \\
574 & Empty/Thin [8] & 12.08 & 0.20 & 0.14 & 0.11 & 0.10 & 0.12 & 0.90 & $<$3.60 \\
599 & Empty/Thin [8] & 12.90 & 0.00 & 0.04 & 0.02 & 0.03 & 0.15 & 0.74 & \nodata \\
642 & Empty/Thin [8] & 12.09 & 0.12 & 0.07 & \nodata & 0.10 & \nodata & 1.07 & $<$1.89 \\
652 & Empty/Thick [24] & 11.56 & 0.02 & -0.01 & \nodata & 0.09 & \nodata & 0.41 & 5.02 \\
682 & Thin/Thick & 11.81 & 0.19 & 0.43 & 1.77 & 2.36 & 3.11 & 4.58 & 7.09 \\
721 & Empty/Thin [8] & 12.64 & 0.05 & 0.01 & \nodata & 0.00 & \nodata & 1.44 & $<$6.23 \\
728 & Empty/Thin [8] & 12.89 & 0.09 & 0.04 & \nodata & 0.10 & \nodata & 0.74 & $<$5.52 \\
729 & Empty/Thin [8] & 12.31 & 0.00 & -0.04 & -0.06 & -0.05 & 0.02 & 0.91 & $<$4.54 \\
740 & Thick (AeBe) & 12.16 & 0.47 & 1.11 & 2.43 & 2.98 & 3.70 & 4.84 & 7.73 \\
826 & Empty/Thin [8] & 13.10 & 0.14 & 0.23 & \nodata & 0.33 & \nodata & 1.27 & $<$4.94 \\
856 & Empty/Thin [8] & 12.64 & 0.05 & 0.01 & \nodata & 0.00 & \nodata & 1.44 & $<$5.20 \\
882 & Thin/Thick & 11.98 & 0.18 & 0.21 & 0.44 & 0.78 & 1.39 & 2.35 & 5.16 \\
901 & Empty/Thin [8] & 12.82 & 0.08 & 0.04 & 0.06 & 0.11 & 0.31 & 0.86 & $<$4.08 \\
902 & Empty/Thin [8] & 12.74 & 0.16 & 0.11 & 0.05 & -0.08 & 0.40 & 1.36 & $<$5.00 \\
912 & Empty/Thin [8] & 12.93 & -0.02 & -0.09 & -0.07 & -0.10 & 0.34 & 1.48 & $<$4.23 \\
917 & Empty/Thin [8] & 13.11 & 0.00 & -0.04 & 0.02 & 0.02 & 0.31 & 1.16 & $<$7.35 \\
923 & Empty/Thin [8] & 12.71 & 0.40 & 0.40 & 0.42 & 0.35 & 0.62 & 1.28 & $<$5.79 \\
\enddata
\end{deluxetable}

\begin{deluxetable}{rlrrrrrrrr}
\tablecaption{IC 1805 Members with Infrared Excesses: Infrared
Excesses Relative to the Photosphere \label{tab:irx-photosphere}}
\tabletypesize{\tiny}
\rotate
\tablewidth{0pt}
\tablehead{
\colhead{(1)} & \colhead{(2)} & \colhead{(3)} & \colhead{(4)} &
\colhead{(5)} & \colhead{(6)} & \colhead{(7)} & \colhead{(8)} & 
\colhead{(9)} & \colhead{(10)} \\
\colhead{opt. num.} & \colhead{Excess Type} & \colhead{\teff} & 
\colhead{$J-H$} & \colhead{$J-K_s$} & \colhead{$J-[3.6]$} & 
\colhead{$J-[4.5]$} & \colhead{$J-[5.8]$} & \colhead{$J-[8]$} & 
\colhead{$J-[24]$} \\
 & & \colhead{(K)} & \colhead{(mag)}  & \colhead{(mag)} & 
\colhead{(mag)} & \colhead{(mag)} & \colhead{(mag)}& \colhead{(mag)}
& \colhead{(mag)}
}
\startdata
        4  &Thin      &11200 &    -0.04  &   -0.06  &    0.16   &   0.26  &    0.38 &     0.65  &    1.52\\
        6  &Empty/Thin [8]       &11200 &     0.04  &    0.02  &       0   &  -0.01  &    0.27 &     0.76  &   $<$5.83\\
       13  &Empty/Thin [8]       &25450 &    -0.05  &    -0.1  &   -0.08   &  -0.05  &    0.09 &     1.03  &   $<$8.15\\
       14  &Empty/Thin [8]       & 9000 &     0.21  &    0.12  &    0.03   &  -0.03  &    0.08 &     0.69  &   $<$5.85\\
       15  &Empty/Thin [8]       &11200 &    -0.13  &   -0.13  &    -0.1   &  -0.03  &    0.02 &     0.74  &   $<$9.24\\
       18  &Thick (AeBe)      &15200 &     0.78  &    1.49  &   -1.05   &   3.21  &    3.76 &      4.7  &    8.40\\
       51  &Empty/Thin [8]       & 9000 &     0.15  &    0.04  &   -0.05   &  -0.13  &       0 &     0.69  &   $<$5.58\\
       61  &Empty/Thin [8]       & 9000 &     0.08  &   -0.09  &   -0.19   &  -0.21  &    0.11 &     0.82  &   $<$7.28\\
       65  &Empty/Thin [5.8]     & 8180 &     0.01  &   -0.01  &    0.12   &   0.21  &    0.56 &     1.52  &   $<$6.40\\
       66  &Empty/Thin [5.8]      & 9000 &     0.08  &    0.02  &    0.06   &   0.01  &    0.82 &     2.32  &   $<$8.39\\
       73  &Empty/Thin [8]       & 9000 &     0.17  &    0.19  &    0.17   &   0.07  &    0.44 &     1.25  &   $<$7.65\\
       80  &Empty/Thin [8]        & 9000 &     0.16  &    0.07  &    0.04   &      0  &     0.1 &     0.98  &   $<$4.88\\
       85  &Empty/Thin [8]       &11200 &    -0.04  &   -0.06  &   -0.05   &  -0.09  &    0.14 &     0.87  &   $<$6.46\\
       86  &Empty/Thin [8]       & 9790 &     0.02  &   -0.03  &   -0.09   &  -0.12  &    -0.1 &     0.57  &   $<$4.80\\
       93  &Empty/Thin [8]       & 8340 &        0  &   -0.04  &   -0.11   &  -0.18  &    0.09 &     0.81  &   $<$5.12\\
       98  &Empty/Thin [8]       & 9000 &     0.03  &    -0.1  &   -0.11   &  -0.12  &    0.17 &     0.85  &   $<$6.86\\
      103  &Empty/Thin [8]       & 9000 &    -0.04  &   -0.15  &   -0.13   &  -0.16  &    0.19 &     1.22  &   $<$7.29\\
      107  &Thick (AeBe)      & 9000 &     0.26  &    0.74  &     1.9   &    2.4  &    2.76 &      3.3  &    4.97\\
      122  &Thin      & 9000 &     0.44  &    0.73  &    1.47   &   1.71  &    1.99 &     3.03  &    3.77\\
      128  &Thick (AeBe)      & 8180 &     0.24  &    0.55  &    1.49   &   1.94  &    2.34 &     3.08  &    5.89\\
      137  &Thick (AeBe)      & 8920 &     0.22  &    0.64  &    1.93   &   2.56  &    3.35 &     4.81  &    7.36\\
      139  &Empty/Thin [5.8]     &16800 &     0.18  &    0.14  &    0.26   &   0.33  &    0.79 &     2.16  &   $<$7.42\\
      143  &Empty/Thin [8]       & 6800 &     0.11  &    0.07  &    0.06   &   0.14  &    0.22 &     1.01  &   $<$7.24\\
      144  &Empthy/Thin [24] &30000 &    -0.01  &   -0.03  &     \nodata    &  -0.03  &   -0.01 &     0.12  &    1.47\\
      180  &Empty/Thin [8]       &11200 &    -0.04  &   -0.12  &   -0.16   &  -0.17  &    0.03 &     0.82  &   $<$5.49\\
      183  &Thin      &34000 &     0.18  &    0.39  &    0.82   &   \nodata    &    1.29 &      1.7  &    2.96\\
      184  &Empty/Thin [8]       &11200 &     0.12  &    0.09  &    0.09   &   0.07  &    0.22 &     0.83  &   $<$5.65\\
      201  &Empty/Thin [8]       & 9560 &     0.19  &    0.16  &    0.23   &   0.17  &    0.43 &     0.83  &   $<$4.82\\
      244  &Empty/Thick [24] &20100 &    -0.09  &    -0.1  &   -0.12   &  -0.14  &   -0.11 &    -0.07  &    3.63\\
      259  &Empty/Thin [8]       & 9000 &      0.1  &    0.05  &   -0.02   &   -0.1  &    0.07 &     0.58  &   $<$6.44\\
      261  &Empty/Thin [8]       & 9000 &     0.09  &   -0.02  &    0.12   &   0.09  &    0.47 &     1.82  &   $<$5.42\\
      270  &Empty/Thin [8]       & 9000 &      0.1  &   -0.01  &   -0.04   &   -0.1  &    0.29 &     1.23  &   $<$4.98\\
      271  &Thin      & 6350 &     0.16  &    0.34  &    0.78   &   1.06  &    1.37 &     1.85  &   $<$0.00\\
      272  &Empty/Thin [8]       & 6650 &     0.01  &   -0.06  &   -0.14   &  -0.23  &    0.02 &     0.61  &   $<$7.37\\
      275  &Empty/Thin [8]        & 9000 &     0.03  &   -0.06  &   -0.09   &  -0.08  &    0.11 &     0.69  &   $<$6.72\\
      279  &Empty/Thin [8]       & 9000 &     0.05  &   -0.02  &   -0.06   &  -0.06  &    0.19 &     0.99  &   $<$4.34\\
      294  &Empty/Thin [8]       & 9000 &        0  &   -0.13  &   -0.19   &  -0.27  &   -0.07 &     0.86  &   $<$4.91\\
      334  &Empty/Thin [8]       & 9000 &    -0.07  &   -0.07  &   -0.07   &  -0.06  &    0.12 &     0.61  &   $<$4.14\\
      340  &Empty/Thin [8]       &11200 &     0.01  &   -0.01  &    0.05   &   0.09  &    0.15 &     0.54  &    3.24\\
      343  &Thick (AeBe)      & 9790 &     0.36  &    0.96  &    \nodata     &   2.38  &     \nodata  &     4.44  &    8.02\\
      386  &Empty/Thin [8]       &11200 &    -0.11  &   -0.13  &   -0.23   &  -0.27  &   -0.23 &     1.51  &   $<$0.11\\
      391  &Empty/Thick [24] & 9000 &    -0.13  &   -0.21  &   -0.28   &  -0.34  &   -0.33 &    -0.15  &    5.65\\
      461  &Thin      & 9000 &     0.15  &     0.2  &    0.55   &   0.85  &    1.08 &     1.63  &     \nodata \\
      520  &Thick (AeBe)      & 9450 &     0.18  &    0.55  &     \nodata    &   2.38  &     \nodata  &     3.45  &    5.22\\
      536  &Empty/Thin [8]       & 9000 &     0.17  &    0.08  &     \nodata    &  -0.04  &     \nodata  &     0.62  &   $<$9.04\\
      563  &Empty/Thick [24] & 9000 &    -0.07  &   -0.17  &     \nodata    &  -0.21  &     \nodata  &     0.31  &    4.01\\
      574  &Empty/Thin [8]       & 9000 &     0.17  &    0.05  &   -0.03   &  -0.07  &   -0.06 &     0.71  &   $<$3.38\\
      599  &Empty/Thin [8]       & 9000 &    -0.03  &   -0.05  &   -0.12   &  -0.14  &   -0.03 &     0.55  &     \nodata \\
      642  &Empty/Thin [8]       & 7500 &     0.04  &    -0.1  &     \nodata    &  -0.21  &    \nodata   &     0.73  &   $<$1.51\\
      652  &Empthy/Thick [24]      &11200 &     0.03  &   -0.02  &     \nodata    &   0.04  &    \nodata   &     0.36  &    4.96\\
      682  &Thin/Thick& 8750 &     0.15  &    0.33  &    1.61   &   2.16  &    2.91 &     4.37  &    6.85\\
      721  &Empty/Thin [8]       & 9000 &     0.02  &   -0.08  &    \nodata     &  -0.17  &    \nodata   &     1.25  &   $<$6.01\\
      728  &Empty/Thin [8]       & 9000 &     0.06  &   -0.05  &    \nodata     &  -0.07  &     \nodata  &     0.55  &   $<$5.30\\
      729  &Empty/Thin [8]       & 9000 &    -0.03  &   -0.13  &    -0.2   &  -0.22  &   -0.16 &     0.72  &   $<$4.32\\
      740  &Thick (AeBe)      & 8340 &     0.41  &    0.98  &    2.24   &   2.76  &    3.46 &      4.6  &    7.45\\
      826  &Empty/Thin [8]       & 8510 &     0.09  &    0.11  &    \nodata     &   0.11  &    \nodata   &     1.04  &   $<$4.68\\
      856  &Empty/Thin [8]       & 9000 &     0.02  &   -0.08  &     \nodata    &  -0.17  &    \nodata   &     1.25  &   $<$4.98\\
      882  &Thin/Thick& 9000 &     0.15  &    0.12  &     0.3   &   0.61  &    1.21 &     2.16  &    4.94\\
      901  &Empty/Thin [8]       & 9000 &     0.05  &   -0.05  &   -0.08   &  -0.06  &    0.13 &     0.67  &   $<$3.86\\
      902  &Empty/Thin [8]       & 9000 &     0.13  &    0.02  &   -0.09   &  -0.25  &    0.22 &     1.17  &   $<$4.78\\
      912  &Empty/Thin [8]       &11200 &    -0.01  &    -0.1  &    -0.1   &  -0.15  &    0.29 &     1.43  &   $<$4.17\\
      917  &Empty/Thin [8]       & 9000 &    -0.03  &   -0.13  &   -0.12   &  -0.15  &    0.13 &     0.97  &   $<$7.13\\
      923  &Empty/Thin [8]       & 9000 &     0.37  &    0.31  &    0.28   &   0.18  &    0.44 &     1.09  &   $<$5.57\\
\enddata
\end{deluxetable}
\clearpage

\begin{figure*}[tbp]
\epsscale{0.75}
\plotone{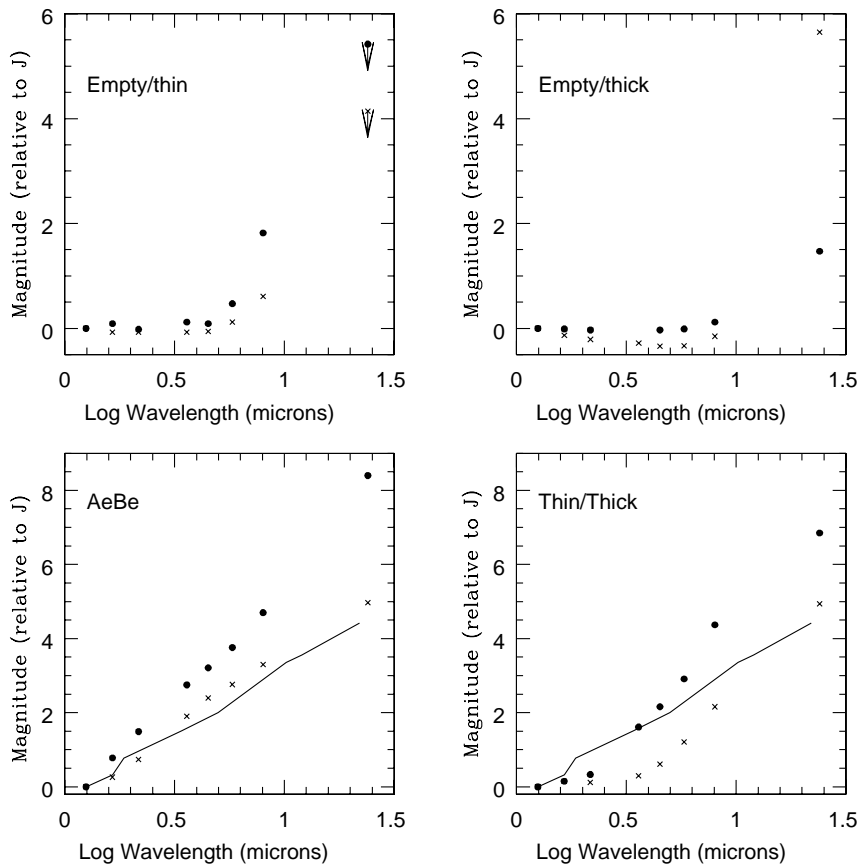}
\caption{A plot showing four of the types of SEDs observed in IC
1805.  (NB: SEDs for stars with optically thin emission are shown in
Figure~\ref{fig:optthinemiss}, and all SEDs appear in the Appendix.)
The magnitudes relative to $J$ have been corrected for reddening, and
the photospheric emission has been subtracted.  In each panel we show
the extremes of the SEDs observed for that type.  Upper left:
Magnitude differences for ``Empty/Thin'' stars, e.g., those  that show
an excess at [8] or in a few cases [5.8], but not at shorter
wavelengths. Filled circles represent the observations of Star No.\
261; crosses, Star No.\ 334. Upper right: Crosses show the SED for
Star No.\ 391, a  typical member of the ``empty/thick" category, i.e.,
stars  that have optically thick emission at [24] but no excess  at
shorter wavelengths.  For comparison, the filled  circles show the SED
for the one star that has optically  thin emission at [24] but no
excess at shorter IR  wavelengths.  Lower left: The SEDs for the stars
with optically thick excesses (Herbig AeBe stars).  Filled circles
represent the observations of Star No.\ 18; crosses, Star No.\ 107.
The solid line in the two lower panels represents the infrared
excesses calculated for a flat reprocessing disk around an A0 star
(Hillenbrand et al.\ 1992).  The calculated excess emission increases
with increasing stellar temperature, and Star No.\ 18 is substantially
hotter than A0 (see Table~\ref{tab:irx-photosphere}).  Lower right:
Stars with optically thin emission in the near infrared but optically
thick emission at longer wavelengths (Thin/Thick).  Filled circles 
represent the observations of Star No.\ 682; crosses, Star No.\ 882. }
\label{fig:irx}
\end{figure*}

\subsection{Classification of Spectral Energy Distributions} 
\label{sec:SED}

After examination of the SEDs of the stars in IC 1805 with excesses
above the photosphere, we find that the SED morphologies can be sorted
into  five categories.  The categories are described below and listed
in Column 2 of Tables~\ref{tab:irx} and \ref{tab:irx-photosphere}. 
The magnitudes of the IR excess relative to $J$ for typical examples
of four of the categories are plotted in Figure~\ref{fig:irx}.  The
stars with optically thin excess emission are discussed in more detail
in Section 5 and their SEDs are shown in Figure 10.   SEDs for all of
the stars with IR  excesses are shown in Figure 11 in the Appendix.

We first list the categories, and then describe them in subsections.
The categories are :
\begin{itemize}
\item Disks with excess emission at $H$ and all longer
wavelengths consistent with that expected for optically thick disks
(``optically thick'').  These could also be described as Herbig AeBe
stars.
\item Disks with optically thin excess emission at $H$ and at
all longer wavelengths for  which a detection was made
(``optically thin'').
\item Disks with optically thin emission at the shorter IR
wavelengths and optically thick  emission at longer wavelengths
(``thin/thick'').
\item Stars for which optically thin excess emission first
appears in either the IRAC bands,  usually only at 8 \mum, or at 24
\mum\ (``empty/thin'').
\item Disks with optically thick emission at 24 \mum\ but
flux densities consistent with  photospheric emission only in $JHK_s$ and the
IRAC bands (``empty/thick'').
\end{itemize}
For the last two categories, the Spitzer wavelength at 
which the excess emission is first detected is listed in 
brackets in the second column of Tables 6 and 7.

\begin{figure*}[tbp]
\epsscale{0.75}
\plotone{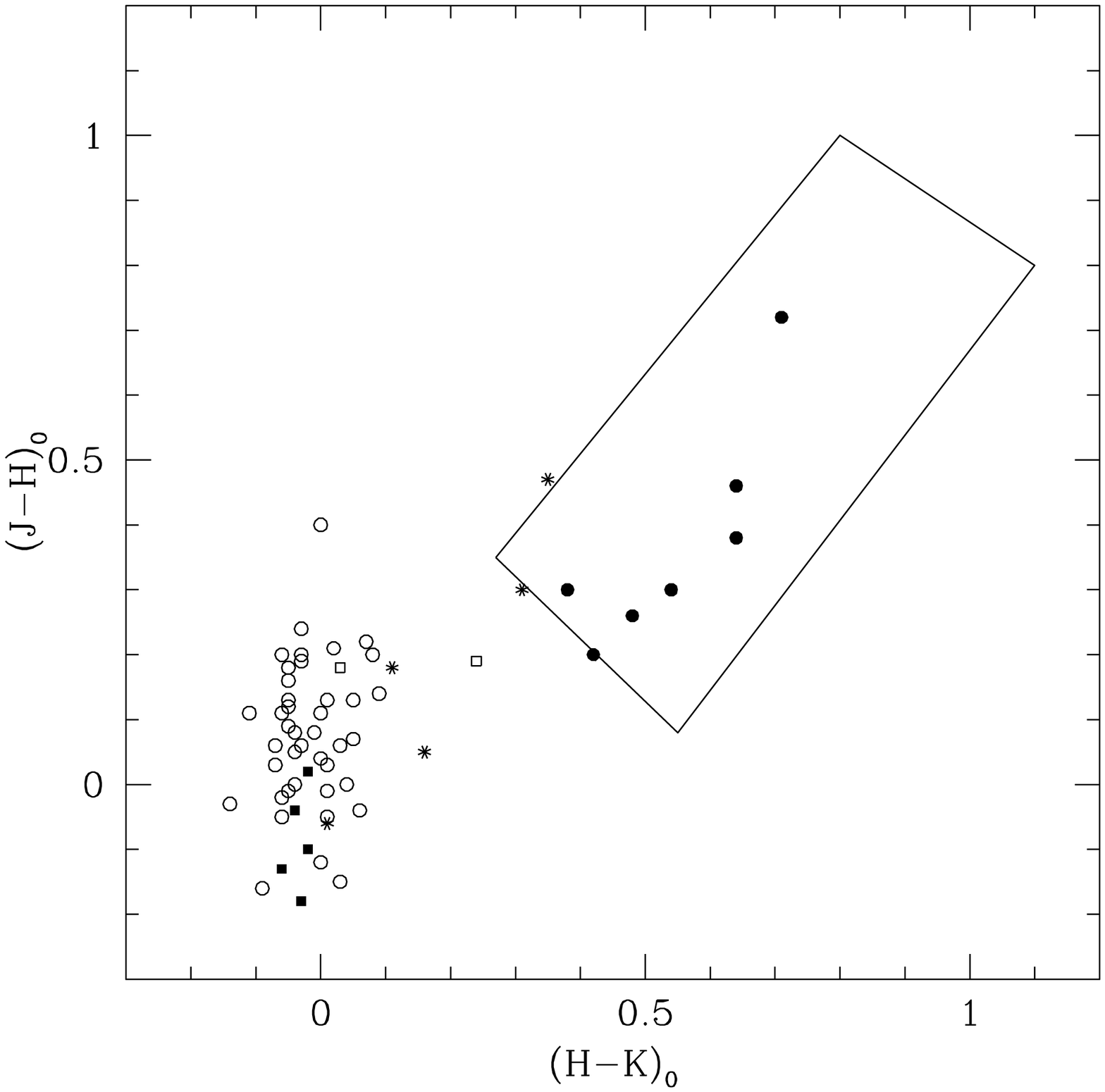}
\caption{ Positions of IC 1805 stars with infrared excesses in the
$JHK_s$ color-color  diagram.  The colors have been corrected for
reddening.  The rectangular region  indicated by the solid lines
shows the area in the diagram occupied by Herbig AeBe  stars
(Hernandez \etal\ 2005).  Filled circles designate ``optically
thick'' (AeBe) stars; asterisks, stars with optically thin emission;
open squares, stars with optically thin emission in the near 
infrared and optically thick emission at longer wavelengths
(``thin/thick''); filled squares, stars with  excess emission only at
[24] (``empty/thick''); and open circles, stars with optically thin
excess emission only in  the IRAC bands, and usually only at 8 \mum\
(``empty/thin'').  The two stars with optically thin  emission that
lie near the bounding box for AeBe stars are Nos.\ 122 and 271.  At
longer  wavelengths, the excess emission for these stars is smaller
than for the AeBe stars in  our sample but for No.\ 122 is somewhat
larger than is typical for the gas decretion disks of classical Be
stars (see Figure 10).}
\label{fig:posofirx}
\end{figure*}

\subsubsection{Optically Thick (AeBe Stars)}

The category of ``optically thick'' is defined as disks with excess
emission at $H$ and all longer wavelengths consistent with that 
expected for optically thick disks. In a recent study, Hernandez
\etal\ (2005) defined  the region in the reddening-corrected
$(J-H)_0$ vs.\ $(H-K_s)_0$ diagram occupied by Herbig  AeBe stars. 
Figure~\ref{fig:posofirx} shows this region along with the
reddening-corrected colors of  the stars in our current sample. Nine
stars lie inside or very near the boundary of the  region defined by
Hernandez \etal\ (2005) and based on their $JHK_s$ colors alone would
be  classified as AeBe stars.  Near-infrared colors do not tell the
whole story, however.   Stars 271 and 122, which lie very near the
AeBe boundary and which would  probably be called AeBe stars based on
$JHK_s$ photometry, have smaller excesses in  the IRAC bands (see
Figure~\ref{fig:ji2i3i4} below) than more typical AeBe stars.   We
have  classified both as optically thin based on the longer
wavelength data, thereby  leaving only seven optically thick (AeBe)
stars.  These stars are labeled ``Thick (AeBe)'' in column 2 of
Tables~\ref{tab:irx} and \ref{tab:irx-photosphere}.

It is generally assumed, based on extensive studies of lower mass PMS
stars that  exhibit similar IR excesses, that objects with SEDs
consistent with emission from an  optically thick disk are also
accreting (e.g., Lada \etal\ 2006). To date, there has not  been a
search for evidence of accretion signatures among a large sample of 
intermediate mass stars selected to have optically thick disks based
solely on their  observed IR SEDs. However, in subsequent discussion,
we make the assumption  that, by analogy with their lower mass
counterparts, intermediate mass objects with  optically thick disks
are likely still accreting material through their circumstellar
disks.

\subsubsection{Optically Thin}

This ``optically thin'' category is defined as disks with optically
thin excess emission at $H$ and at all longer wavelengths for  which
a detection was made.  This group is far from homogeneous.  Two stars
(122 and 271) have $JHK_s$ excess nearly as large as the optically
thick (AeBe) stars but lower excesses at  longer wavelengths. 
Surprisingly, the  upper limit on the excess at 24 \mum\ for No.\ 271
is 0 mag.   One star in this category, No.\ 461, may be blended with
No.\ 292.  Another star, No.\ 183,  was classified as a classical Be
star by Massey \etal\ (1995a).  Classical Be (CBe) stars  are losing
mass, and their IR excesses are produced by free-free emission
arising in a  circumstellar decretion disk. CBe stars occupy a region
in the $JHK_s$ two-color plot near  the origin; 90\% have excesses in
both $(J-H)_0$ and $(H-K)_0$ that are $<$0.3 mag.  (Dougherty,
Taylor, \& Clark 1991; Dougherty \etal\ 1994).  All five of these
stars are labeled ``Thin'' in column 2 of Tables~\ref{tab:irx} and
\ref{tab:irx-photosphere}, and their SEDs are shown in Figure 11. 

\subsubsection{Thin/Thick}

This category (``Thin/Thick'') is defined as disks with optically
thin emission at the shorter IR wavelengths and optically thick 
emission at longer wavelengths.  This category includes only 2 stars,
Nos.\ 682 and 882.   Star 682 has smaller excesses at $H$ and $K_s$
than is typical of optically thick (AeBe) stars, but the excess in
the IRAC bands and at 24 \mum\ is similar to that of optically thick
(AeBe) stars.  Star No.\ 882 has less excess emission than typical
AeBe stars out through the IRAC bands but optically thick emission
at 24 \mum.  These stars are labeled Thin/Thick in Column 2 of
Tables~\ref{tab:irx} and \ref{tab:irx-photosphere}.

\subsubsection{Empty/Thin}

This ``Empty/Thin'' category includes stars for which optically thin
excess emission first appears in either the IRAC bands,  usually only
at 8 \mum, or in one case only (No.\ 144) at 24 \mum.  The
wavelength  at which the excess first appears is listed in Column 2
of Tables~\ref{tab:irx} and \ref{tab:irx-photosphere} (``Empty/Thin
[5.8]'', [8], or [24]).  These are disks whose inner regions are
devoid of small dust grains and/or emitting gas within the limits of
our ability  to assess excess emission exceeding the uncertainties in
our photometry and/or our  knowledge of reddening. 

Three of these stars, Nos.\ 65, 66, and 139, first exhibit excess
emission at [5.8], and the  excess is larger at [8], suggesting that
we are indeed seeing emission from a disk in  which the inner region
as been cleared of small dust grains.

For the remaining 42 stars in this category, the abrupt SED rise 
between 5.8 \mum\  and 8 \mum\ is too large to be explained easily by
a continuous energy distribution  such as those seen in either the
stars with optically thin emission or in the optically thick (AeBe) 
stars.  The lack of a short wavelength ``Wien tail'' to the SED of
the excess emission  observed at 8 \mum\ suggests that the excess at
8 \mum\ may be caused by polycyclic aromatic hydrocarbon (PAH)
emission  rather than thermal emission from interstellar-like dust
grains.  Indeed, these stars  occupy a position in a plot of
[3.6]$-$[5.8] vs.\ [4.5]$-$[8] similar to that of PAH-rich galaxies
(see Figure 9).   (See, e.g., Hernandez \etal\ 2007 for a comparison
of colors of galaxies and pre-main sequence stars.) The IC 1805
objects, however, have stellar PSFs and to our limiting  magnitude we
expect negligible contamination from galaxies (Fazio \etal\ 2004b).  
Moreover, we have spectral types for 16 of these objects and can thus
confirm that the  excess emission arises from optically-identified
members of the IC 1805 complex.

The question then is whether the emission is an artifact of the data
reduction process or,  if real, whether it arises arises from PAH
background emission or from the stellar disk.   In order to address
this question, we conducted several different tests. 

First, we manually investigated the images of each object,
particularly those at 8 \mum.  Some of the objects with excesses only
at [8] are located in regions where there is no bright 8 \mum\  ISM
emission. 

Next, we tried different reduction procedures.  The IRAC photometry
extraction  described above in \S\ref{sec:obs} used an aperture
of 3 px and a sky annulus of 3 to 7 px. We also  performed the 8 \mum\
photometry using different settings (e.g., 2 px and 2 to 6 px, with 
the associated aperture correction), and the same measurement was
obtained to within  0.3 mag (usually well within that) and below the
0.5 mag limit we set above for  identifying an 8 \mum\ excess. 

If the flux density measured using the same aperture/annulus at a
nearby location  was comparable to the measured target flux density,
then the measurement might well be  spurious. We took our 3 px
aperture/3-7 px annulus measurements and performed the same photometry
offset by 8 px in the $x$ direction in the 8 \mum\ images, which is
close to but not exactly a shift  in RA. This offset was performed
blindly, e.g., no effort was made to ensure that there  were no
adjacent sources at that location. For the overwhelming majority of
our sources  in general and all but 4 of the sources identified as 8
\mum\ excess sources, this offset  flux density was more than 0.3 mag
fainter, often significantly (more than 3 mag) fainter than the  flux
density measured for the nearby 8 \mum\ source. 

If the flux density measured in the sky annulus were to exceed that
measured in the  target aperture, then the derived 8 \mum\ flux
density could well be spurious. For the  overwhelming majority of our
sources in general and all but 1 of the sources identified  as 8 \mum\
excess sources, the object flux density significantly exceeds that
measured in the  sky annulus.

Finally, we looked at the distribution of sky annulus flux density
values to see if the  objects identified as 8 \mum\ excess objects
fell only in regions where  this value is the highest. The 8 \mum\
excess objects are not preferentially located in  regions where the
measured sky annulus flux density is highest. There are even a 
significant number of objects with no 8 \mum\ excess and with
significantly higher sky  annulus flux densities than the highest
found in the 8 \mum\ excess sample. 

We conclude that the 8 \mum\ excesses are real, likely arise from PAH
emission, and are  most probably associated with a circumstellar
disk.  Diagnosing the  properties of such putative disks would require
measurements at $\geq$24 \mum.  Unfortunately, all but four of the
stars in this category have upper limits  for \ks$-$[24] that are
larger than 4 mag, and so the disks could be either  optically thick
or optically thin (see, e.g., Hernandez \etal\ 2007 for a sample of
\ks$-$[24]  magnitudes for pre-main sequence stars).  One of the
stars, Nos.\ 386, has an upper limit  for \ks$-$[24] of 0.11, which is
consistent with no excess.   Nos.\ 574 and 642 have upper limits for
\ks$-$[24] of 3.38 and 1.51, respectively, and No.\ 340 has a measured
value of 3.24.  All three are consistent with excesses measured for
debris disks; for  comparison, \ks$-$[24] for $\beta$ Pic, which has
one of the largest 24 \mum\ excesses known, is 3.5 (Rebull \etal\ 2008).

\subsubsection{Empty/Thick}

This last category consists of disks with optically thick emission at
24 \mum\ but flux densities consistent with just photospheric
emission in $JHK_s$ and the IRAC bands.  The four stars in this
category are designated as ``Empty/Thick'' in Column 2 of
Tables~\ref{tab:irx} and \ref{tab:irx-photosphere}.  Owing to the 
high  and variable upper limits to 24 \mum\ flux densities listed in
Table~\ref{tab:spitzerfluxes}, the number of stars in this  category
is almost certainly a lower limit to the true frequency of such
objects.  

\subsubsection{Discussion of categories}

\begin{figure*}[tbp]
\epsscale{0.75}
\plotone{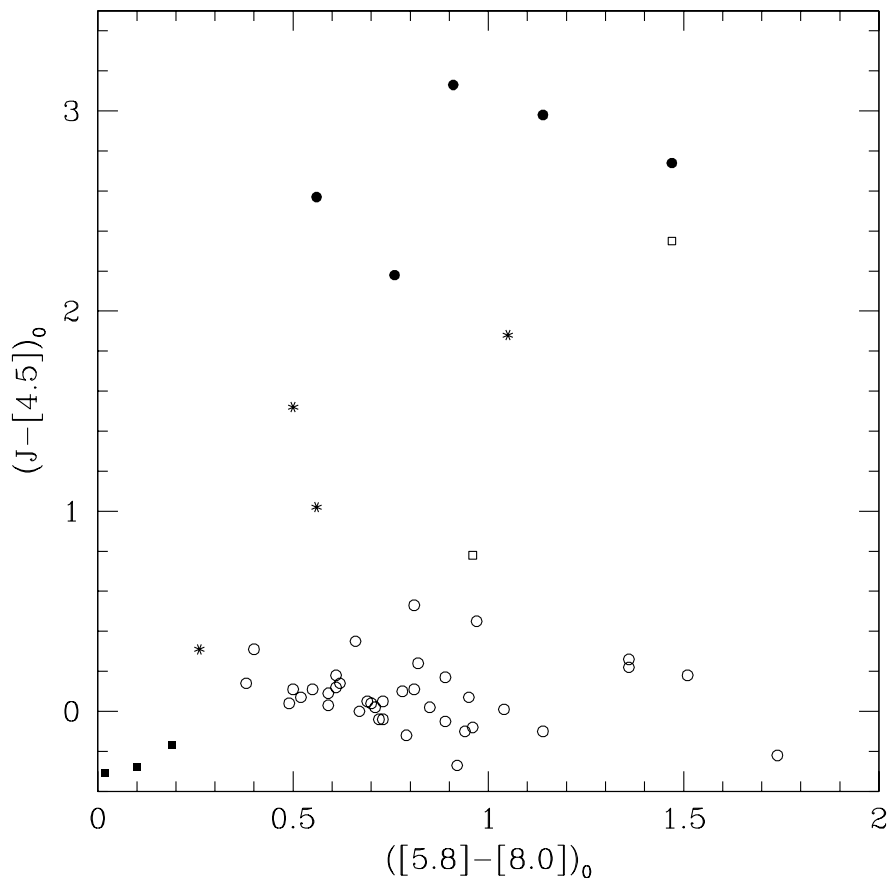}
\caption{ A plot of $(J-[4.5])_0$ vs.\ $([5.8]-[8.0])_0$ (i.e., color
corrected for reddening but  uncorrected for photospheric emission). 
The symbols are the same as in Figure~8 (filled circles=optically
thick (AeBe);  asterisks=optically thin; open circles= empty/thin; 
open squares= thin/thick),  but not all of the stars detected in
$JHK_s$ were detected in all three of the IRAC bands used in this
figure.  The optically thick (AeBe) stars are clearly separated from
the rest of the sample.  We also see that the infrared excesses
relative to $J$ for Nos.\ 122 and 271 (the two optically  thin stars
with the largest values of $(J-[4.5])_0$) are lower than  for the
optically thick (AeBe) stars.  Note also the large number of stars
(empty/thin, designated with open circles)  with excesses between
0.75 and 1.5 mags in ([5.8]$-$[8.0]) but no excess at shorter 
wavelengths.}
\label{fig:ji2i3i4}
\end{figure*}

As Figure~\ref{fig:posofirx} shows, most of the stars in this sample
with IR excesses would not be identified with $JHK_s$ photometry
alone.  Two of the stars just outside the boundary of the AeBe star
region (Nos.\ 271 and 122) would probably be classified as  AeBe
stars based on their near infrared  photometry alone.  These stars,
however, have lower excesses in the IRAC bands than the optically
thick (AeBe) stars.  This is illustrated in Figure~\ref{fig:ji2i3i4},
which shows a plot of  $(J-[3.6])_0$ vs.\ $([5.8]-[8])_0$.  In this
plot, the colors have been corrected for reddening but the
photospheric flux density has  not been subtracted.  The optically
thick (AeBe) stars are well separated from the other categories by 
virtue of their large excesses at shorter wavelengths.  This diagram
also shows the  large number of stars for which we measure excesses
of 0.5-1.75 mag in $[5.8]-[8]$ but  no excesses at shorter
wavelengths (empty/thin stars).  This region is  also well-populated
in the same  color-color plot for W5 (see Koenig \& Allen 2010). 

The current sample was optically selected, and one might be concerned
that the stars  with disks are on average more reddened (embedded)
than those without, and that the  effective magnitude limit of the IR
excess stars is brighter than for the no-excess stars.   The averaged
extinction in $V$ is, however, similar for the two groups:  \av=2.50 for
the  cluster members with excesses and 2.67 for the cluster members
with no IR excess.   This sample should, therefore, allow us to
compare directly stars that are in various  stages of disk evolution
subsequent to the deeply embedded phase.

\subsection{Percentage of disks around B and A type stars}
\label{sec:statmembers}

With these data, we can estimate what percentage of stars more massive
than 2 \msun\ and less than 3 Myr old have disks.  Obviously,
we know the fraction of stars  with IR excesses among the confirmed
members.  However, membership for the cooler  stars was based largely
on spectral classification, and stars with IR excesses are  probably
over-represented in the spectroscopic sample; when we constructed the
list of  stars to be observed spectroscopically, we had already
partially reduced the Spitzer  data and had identified the stars with
the largest IR excesses, which were then  given high priority
for spectroscopy. For example, all of the stars subsequently
classified  as AeBe stars were included in the spectroscopic sample. 

By contrast, the stars without IR excesses chosen for spectroscopy
were selected at  random from stars observed by Massey \etal\ (1995a).
In order to estimate the true  fraction of stars that show IR
excesses, we have used the percentage of cluster  members with no IR
excess found among our spectroscopic sample to estimate the 
percentage of likely cluster  members among the stars for which we
only have  photometry.  To do so, we divided the stars with
spectroscopy and no IR excess into  bins according to observed $V$
magnitude and counted the fraction of stars in each bin  that were
determined to be members based on spectral classification.  We then
similarly  divided the sample of stars with photometry and no
membership information into  magnitude bins and applied the correction
factors derived from the spectroscopic  sample to estimate the likely
number of members.  On average, 75\% of the no-excess  stars with
spectra proved to be members, with some evidence for increasing 
contamination by non-members at fainter magnitudes.  Applying the
correction factors to  each magnitude bin for the stars without
spectra, we find that 288 of the 391 stars  without spectra are likely
to be members. 

Table~\ref{tab:membstatus} summarizes the membership of IC 1805
including the basis for deciding  membership.  The first three rows in
this Table include the stars determined to be  members based on their
reddening (i.e., the stars in Table~\ref{tab:members}).  The fourth
row includes  all of the additional stars with IR excesses, and we
have assumed that all stars  with IR excesses are members. If we then
include the likely 288 additional stars,  the total number of cluster
members in our sample is estimated to be 548, of which 63  or only
11.5\% have IR excesses.  

\begin{deluxetable}{lp{11cm}}
\tablecaption{Members of IC 1805\label{tab:membstatus}}
\tablewidth{0pt}
\tablehead{
\colhead{No.\ of stars} & \colhead{Membership Status} }
\startdata
81 & Stars with no excess IR emission but reddening derived by
Q-method consistent  with cluster reddening\\ 
116 & Stars with no excess IR emission but reddening derived from
spectral  classification consistent with cluster  reddening\\
32 & Stars with IR excesses and reddening 
derived from either the Q-method or spectral 
classification consistent with cluster 
reddening \\
31 & Stars assumed to be members because of 
their IR excesses \\
288 & Stars with no IR excess and no information 
about reddening but on a statistical basis 
likely to be members as determined from 
the ratio of members/non-members in the 
spectroscopic sample \\
\hline
548 & Total number members (63 have IR excesses)\\
\enddata
\end{deluxetable}

The numbers of disks of each type are listed in
Table~\ref{tab:sedtype}. The last  column of the Table gives the
percentage of IC 1805 members that have disk  emission in each of the
five categories. Only about 1.3\% of the total sample have  optically
thick disks, i.e., are of the Herbig AeBe type. All of the stars with
optically thick  disks have masses between 2 and 4 \msun, and the
fraction of optically thick disks in  this more restricted mass range
is still only 1.6\%.  This should be compared with low  mass stars,
about half of which retain their optically thick disks (i.e., remain
classical T  Tauri stars) at an age of about 3 Myr (Haisch \etal\
2001).

Although all the same types of SEDs are seen among lower mass stars
(e.g., Lada \etal\ 2006; Muzerolle \etal\ 2006), the proportions are
quite different.  Specifically, the majority  ($\sim$9\%) of the stars
with excesses among our sample of intermediate mass stars have 
excesses only at wavelengths longer than 5 \mum.  A similar result is
found for  intermediate mass stars in W5 (Koenig \& Allen 2010) and
NGC 6611 (Rebull, Wolff, \&  Strom, unpublished).  SEDs of this type
are rare among stars with masses less than that of the Sun. It is 
important to note, when comparing SEDs of intermediate- and low-mass
stars, that the excesses at specific wavelengths probe emission
arising from very different  disk radii.

\begin{deluxetable}{p{8cm}p{6cm}ll}
\tablecaption{Total number of SED types \label{tab:sedtype}}
\rotate
\tablewidth{0pt}
\tablehead{
\colhead{Type of SED} & \colhead{Possible interpretation} & 
\colhead{No.\ of stars} & \colhead{Fraction of cluster members} }
\startdata
Optically Thick (AeBe) & Likely Accretion Disk & 7 & 1.3\% \\
Optically Thin (thin throughout near IR and at  24 \mum\ if
detected) & Either gas emission or homologously depleted disks
&  5 & 0.9\% \\
Thin/Thick (Near IR thin/24 \mum\ thick) & Possible EGP Formation &  2 & 0.4\% \\
Empty/Thin (No excess in $JHK_s$;  optically thin emission first  appears at
$\lambda_{\rm initial}$) & Emission from second generation dust &  45&  8.2\% \\
Empty/Thick (Optically thick emission at 24 \mum; no excess  at shorter
wavelengths) & Photoevaporation or companion formation &  4 & 0.7\% \\
\enddata
\end{deluxetable}
\clearpage

Another difference between the current intermediate mass sample and
disks around  stars with masses similar to or lower than that of the
Sun is the large number of disks that are apparently in an evolutionary
state more advanced than the initial, optically thick accretion phase. 
Of the stars with disks, only 11\% are optically thick
while 89\% appear to have SEDs indicative of a later phase of evolution.  
This contrasts strongly with lower mass stars (Lada \etal\ 2006; 
Muzerolle \etal\ 2010).

\section{Discussion}
\label{sec:discussion}

\subsection{Factors Determining Disk Properties}
\label{sec:diskprops}

The data presented in Tables~\ref{tab:irx} and
\ref{tab:irx-photosphere} provide a starting point for discussing the 
evolution of disks among intermediate mass stars. To understand the different 
types of SEDs observed, we  need to consider both the likely range of disk
initial conditions and the factors that affect  the evolution of
accretion disks. 

{\bf (a) Disk initial conditions.} There is a paucity of information 
available regarding the initial properties (M and \mdot) among
intermediate-mass stars.  Hence, we must use the extensive data
available  for the low-mass stars as a guide.  Andrews \& Williams
(2007) provide a summary of the  frequency distribution of disk masses
among the young ($t <$ 2 Myr) populations in  two nearby star-forming
regions: Taurus and Ophiuchus. Their results, based on  sub-millimeter
and millimeter-continuum estimates of disk masses, show a wide range
($>$ 2 dex)  of masses.  To provide some representative numbers,
Andrews \& Williams (2007)  find that $\sim$50\% of optically thick
disks surrounding stars with masses in the range 0.2 to 1 \msun\ have
$M_{\rm disk}>$0.01 \msun, while only about 10\% of their samples 
contain $M_{\rm disk}>$0.05 \msun. Andrews \& Williams (2005, 2007) 
also find that the upper envelope of the relationship between $M_{\rm
disk}$ and the  mass of the central star ($M_{\rm star}$) scales
linearly with $M_{\rm star}$ (higher disk mass for higher  mass
stars). It is noteworthy that the range in disk accretion rates (e.g.,
Muzerolle \etal\ 2003)  also spans a range exceeding 2 dex. For an
accretion disk modelled as an  ``alpha disk,'' \mdot $\sim \alpha
\times \Sigma$, where $\Sigma$ is the disk surface density, $\sim
M_{\rm disk}/r^2$. For disks with homologous surface density-radius
distributions, \mdot\ $\sim  M_{\rm disk}$. Were these simple
assumptions  applicable, much of the observed range in disk accretion
rate could be attributed to the observed range of disk masses. Of
course, nature is  likely to be more complex, and at least some of
the variation in \mdot\ could arise from differences in the radial
distribution of disk surface density as well as disk viscosity.

{\bf (b) Disk Evolution.} Three factors drive disk evolution: (i)
draining and spreading of an  accretion disk as a result of angular
momentum evolution (e.g.,  Calvet, Hartmann, \&  Strom 2000); (ii)
photoevaporation of the disk driven by a combination of X-ray,
far-ultraviolet, and extreme ultraviolet radiation acting on a
draining and spreading disk (see Gorti,  Dullemond, \& Hollenbach 2009
for a current discussion and review of previous  work); and (iii)
grain settling,  grain growth, and the formation of planetesimals and
planets, which can alter the dust/gas  ratio,  the dust opacity, and
the distribution of gas and dust via formation of tidal gaps.  The
timescale  to deplete the initial disk mass by a factor of 100  via
draining and (consequent  spreading) is $10^8$ yr (e.g., Calvet,
Hartmann, \& Strom 2000).  Based on the Gorti \etal\ (2009)
assumptions, the time for photoevaporation to remove gas and dust from
a  circumstellar disk with initial mass of 0.1 \msun\ surrounding a
star of mass $\sim$2 \msun\  is $\sim$4 Myr; for stars with mass $M >
4$ \msun, the comparable timescales drop rapidly  with increasing mass
(to 0.5 Myr for a 7 \msun\ star). The effects of planet formation  on
the lifetime of accretion disks with initial high masses and rates of
disk destruction are not well constrained at present.

This brief summary of the range of disk initial conditions and the
factors that together affect disk evolution provides a framework for
discussing the observed evolution of disk  properties of intermediate
mass stars at least in a rudimentary way. In the following
subsections, we provide a summary of plausible physical explanations
for each of the  observed SED classes and suggest additional
measurements that in principle should  provide the basis for choosing
among these explanations.

\subsection{SEDs consistent with optically thick accretion disks}

As we argue in the previous section, our data show that at ages
of 1-3 Myr, only 2\%  of stars with $2 < M/$M$_{\odot} < 4$ are still
surrounded by optically thick accretion disks. No  evidence of
optically thick emission is found for stars $M >$ 4 \msun. Studies
(Haisch \etal\ 2001; Lada \etal\ 2006) of disk properties for solar
like stars ($M \lesssim 1$ \msun) show that at $t  \sim$2 Myr, a much larger
fraction of such lower mass objects (30-60\%) exhibit evidence  for
optically thick disks.  (Such disks are sometimes categorized as
``primordial disks,''  despite the fact that although still accreting
and optically thick, the mass and structure of  such accretion disks
must have changed significantly since the star-disk system ceased  to
be fed by infalling material from its natal protostellar core. We
suggest that this  nomenclature be re-examined [e.g., Evans \etal\
2009]).  By comparison, the fraction of  solar-like stars surrounded
by accretion disks does not drop to values as small as 6\%  until
ages $t \sim$ 5-7 Myr (Dahm \& Hillenbrand 2007; Haisch \etal\
2001).  Our results are qualitatively consistent with the trend
reported by Dahm \& Hillenbrand  (2007), who find no evidence for
optically thick accretion disks for stars with $M >$ 1.2 \msun\ in IC
2362 ($t \sim$ 5 Myr) and with earlier results reported by
Hillenbrand \etal\ (1993),  who conclude that no stars with $M >$ 5
\msun\ in the young cluster NGC 6611 show  evidence of optically
thick accretion disks at $t \sim$ 1-2 Myr.

We propose that a combination of photoevaporative erosion of disks
combined with  differences in initial disk masses can explain the
absence of optically thick accretion disks among stars with $M >$ 4
\msun, and also the small fraction of  stars showing evidence of
optically thick disks among stars with $M \sim$ 2 \msun.  As noted
above, Gorti,  Dullemond, \& Hollenbach (2009) estimate the
timescale  for photoevaporation to erode a ``typical" disk ($M\sim
0.05 M_{\rm star}$) surrounding  intermediate mass stars spanning a
range of masses. For a star having $M \sim$ 2 \msun,  they estimate a
disk lifetime of 4 Myr presuming that photoevaporation is the primary
cause for disk dissipation; for $M_{\rm star}\sim$ 7 \msun, they
estimate a disk lifetime $t \sim$ 0.5 Myr. Their results suggest that
the rapid increase in photoevaporation  rate with increasing mass (a
result of the dramatic rise in extreme-ultraviolet (EUV) luminosity
as a  function of mass along the main sequence) provides a natural
explanation for the  paucity of optically thick disks found among
stars with $M >$ 4 \msun\ in a cluster  whose members span ages 1-3
Myr.

The small fraction of optically thick disks ($\sim$2\%) found to
surround  stars of $M\sim$ 2 \msun\ may reflect the fact that the median
disk mass around  intermediate mass stars may be 0.01 $M_{\rm star}$
rather than 0.05 $M_{\rm star}$ (as assumed by Gorti, Dullemond, \&
Hollenbach 2009),  if these objects  have a distribution of $M_{\rm
disk}/M_{\rm star}$  similar to that  characterizing lower mass stars
(Andrews \& Williams 2005, 2007).  If so, then the lifetime of the
median disk  around a star with $M \sim$ 2 \msun\ will be reduced by a
factor of 5 from the 4 Myr  lifetime estimated by Gorti, Dullemond and
Hollenbach (2009) for a disk of  0.05 $M_{\rm star}$. If so, the
lifetime of a median mass disk  for a 2 \msun\ star would be $t \sim$
0.8 Myr. Hence, by an age of $t \sim$ 3 Myr, we would expect all but
the most massive disks among the cohort of 2 \msun\ stars in IC 1805 
to have been eroded by photoevaporation. The 2\% that still possess
disks would  plausibly represent that small fraction of stars with
disks masses near the  expected upper limit of 0.1 $M_{\rm star}$.

\subsection{SEDs consistent with optically thin disk emission}

Disks classified as ``optically thin'' show measurable excess emission
above  photospheric levels (at all IRAC bands, and where available,
MIPS-24), but at levels that  lie at least a magnitude (and more
typically 2-3 magnitudes) below those characteristic  of geometrically
flat, optically thick accretion disks (i.e., disk structures that
would obtain  in an accretion disk in which grains have settled to the
disk midplane at radii $r \lesssim$10-20  AU; see Hillenbrand \etal\
1992). Only five such disks are found among our sample of IC 1805
members, and one of those is likely a CBe star.  This represents 0.9\%
of our total  sample, and so such disks are rare. Below, we discuss
two possible origins for the observed excess emission: from optically
thin gas and optically thin dust.

The possibility that gas emission accounts for the observed excesses
in stars showing SEDs similar to our optically thin class was
discussed by  Hillenbrand \etal\ (1992) in their study of a large
sample of Herbig AeBe stars; the  authors adopted the nomenclature
``Group III'' for these objects.  This nomenclature has  sometimes
resulted in confusion given the by now common practice of referring to
low  mass PMS stars that lack any infrared excess emission as ``Class
III'' objects.  The  objects in  ``Group III'' in the Hillenbrand
\etal\ (1992) sample of AeBe stars show strong H$\alpha$ emission,
have spectral types ranging from B0-B3, and are embedded within
molecular  cloud complexes (as evidenced by their illuminating nearby
reflection nebulae).  Hillenbrand \etal\ (1992) speculated that these
objects might be analogous to classical Be  stars, whose H$\alpha$
emission lines and small IR excesses are believed to arise in  gaseous
decretion disks produced as a consequence of rapid rotation. 

As noted, one optically thin disk found in our study of IC 1805
surrounds a B2 star,  which was classifed as a Be star by Massey 
\etal\ (1995a). One other (No.\ 4) is a B8  star.  We do not have
spectral types for two others (No.\ 122 and 461), but given  their
observed $V$ and $J$ magnitudes, they are also likely to be late B
stars.  No.\ 271,  which is on the borderline between optically thick
and thin emission, is an F8 star.

\begin{figure*}[tbp]
\epsscale{0.75}
\plotone{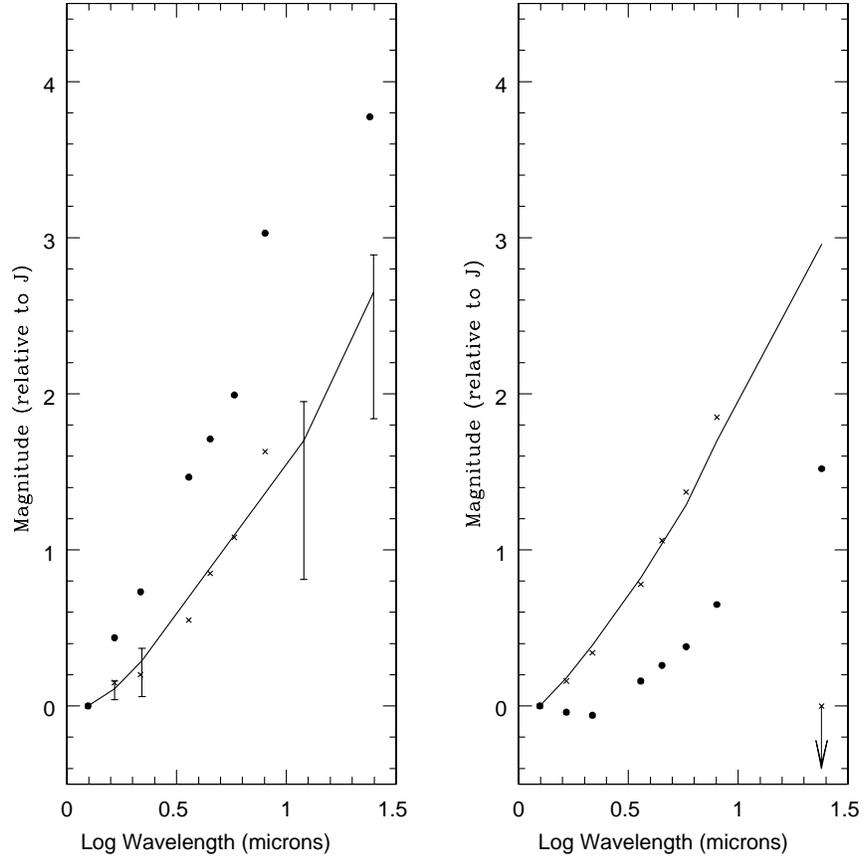}
\caption{Magnitudes relative to $J$ for stars with optically thin
emission.  The magnitudes have been corrected for reddening and the
photospheric  emission has been subtracted.  (Left) The solid line
indicates the median emission  observed for  classical Be stars (see
text for references).  The error bars indicate the emission for the
25th and 75th percentiles for the observed sample.  Filled circles and
crosses indicate the observed magnitude differences for Star Nos.\ 122
and 461, respectively.  (Right)  The solid line represents the
magnitude differences  relative to $J$ observed for Star No.\ 183, which
was classified as a classical Be star  by Massey \etal\ (1995a). 
Filled circles and crosses indicate the observed magnitude differences
for Stars No.\ 4 and 271, respectively.}
\label{fig:optthinemiss}
\end{figure*}

Thus four of the optically thin disks in our sample are likely
to be considerably later  in spectral type than those comprising the
``Group III'' stars discussed by Hillenbrand \etal\ (1992) and also
later in spectral type than the majority of CBe stars.  Unfortunately,
the spectra used to classify stars in our sample lack the spectral 
resolution needed to identify H$\alpha$ emission objects
unambiguously, particularly given  the high and variable emission
arising from the \ion{H}{2} region surrounding the cluster. 
Nevertheless, their SEDs provide the basis for testing whether their
excess emission  \emph{could} arise from a gaseous disk. In
Figure~\ref{fig:optthinemiss}, we plot the SEDs for each of our
optically  thin disks. Superposed on this plot is the median SED,
along with the upper and lower  quartile SEDs derived from
observations of a sample of 101 CBe stars (Cote  and Waters 1987).
This Figure shows that the shape of the SEDs for these objects  mimics
those that characterize the gaseous disks of CBe stars.  

Optically thin emission could also arise from dust emission. Currie
\etal\ (2009) have indeed argued that among low mass PMS stars, such
SEDs arise in  ``homologously depleted disks (HDDs).''  According to
these authors, such disks  represent a relatively long-lived
``transition phase'' as optically thick accretion disks begin  to
develop systems of planetesimals, and later, planets. They suggest
that HDDs are significantly depleted in small dust grains relative to
typical optically thick accretion  disks.  The SED slopes for the
excess emission produced by these disks approximately  follow a power
law through 24 \mum, a result similar to that shown in
Figure~\ref{fig:optthinemiss}. In analogy  to our ``optically thin
disks,'' the disks classified as HDDs are underluminous compared to 
an optically thick, geometrically thin disk by a factor of $>$ 2-3
from 5.8 \mum\ to 24 \mum.  Currie \etal\ suggest that these features
are consistent with a disk that loses a  significant  amount of mass
of small dust grains at all disk radii simultaneously, which  they
attribute  to the growth of dust to larger bodies.

Critical to distinguishing between competing explanations (gas vs.\
dust emission) for  optically thin disks around pre-main sequence stars 
of all masses would be measurements of (a) Balmer
emission line profiles, whose  observation would reveal the likely
presence of circumstellar gas, and whose  morphology would reveal
whether or not the gas is accreting; (b) 
high spectral resolution measurements of CO or other gas tracers; the
shape of such  features would provide incontrovertible evidence of gas
in Keplerian motion around the  star; and c) silicate emission
features, whose presence would provide clear evidence of  heated dust
in an optically thin disk and whose morphology would diagnose whether
or  not significant grain growth has taken place. Until such
measurements are available, it  will be impossible to distinguish
between a gas or dust origin for the excess emission  arising in
optically thin disks around pre-main sequence stars.

\subsection{SEDs consistent with optically thin inner and optically
thick outer disks (thin/thick)}

These sources have weaker $JHK_s$ and IRAC flux densities compared to those
corresponding to  optically thick accretion disks, but 24 \mum\
flux densities that are consistent with emission  arising from optically
thick disks. There are low mass analogs among well studied  examples
of stars with ``inner holes and gaps'' (of size 2-10 AU) such as 
Hen 3,  TW Hya (Calvet \etal\ 2005; Low \etal\ 2005; see also
Strom \etal\ 1989 and Skrutskie \etal\ 1990 for an historical
perspective on this class of disks, initially dubbed ``transition 
disks''). Only two stars (Nos.\ 682 and 882) fall in this category --
that is, 1/4 the number  with optically thick disks and only
$\sim$0.4\% of the total sample of IC 1805 members. Note,  however,
that in most cases, the high upper limits to 24 \mum\ flux densities
mitigate against  firm detection of other such objects which may be
present in our optically-selected  sample of candidate members.

The small excesses found at short IR wavelengths in such disks have
been attributed to  emission arising from small dust grains located in
inner disk regions (in these cases, $r <$  20 AU).  Based on this
hypothesis, Robitaille \etal\  (2006) estimate that $10^{-7}$ to
$10^{-5}$ \msun\ of small dust grains can explain the SEDs for 
``typical'' thin/thick disk surrounding  low mass stars. Mass
estimates based on mm- and  sub-mm measurements suggest outer disk
masses ranging from $10^{-1}$ to $10^{-2}$ \msun\  (Najita, Strom, \&
Muzerolle 2007), but see Marin \etal\ (2010) and Cieza \etal\ (2010)
for examples of  similar objects with much lower disk masses.

Najita \etal\ (2007) have suggested that the observed properties of
thin/thick disk  systems, at least among low mass pre-main sequence
stars, can best be explained by  positing the formation of a
Jovian-mass planet in the inner regions ($r <$ 10 AU) of an 
accretion disk having total mass and surface density consistent with
forming planets of  $M \sim$ 1 $M_J$. Simulations suggest that in
such disks, accretion onto the star will continue  via ``accretion
streams,'' albeit at rates $\sim10\times$ lower than those
characterizing accretion  disks of comparable mass but in which giant
planet formation has not (yet?) taken place  (e.g., Lubow, Seibert,
\& Artymowicz 1999). The Jovian mass planet opens up a gap,  whose
size depends on the mass of the planet and the physical properties of
the  accretion disk. The lower accretion rate onto the star results
because the planet  accretes $\sim$90\% of the material flowing
inward from the outer regions of the (still) optically  thick outer
disk regions. The duration of this phase is expected to be short
($\sim10^5$  years), owing to the rapid rate at which material
accretes onto the Jovian mass planet;  once the planet reaches a
sufficiently high mass, it can truncate the disk tidally, and halt 
accretion onto the star. The presence of optically thin emission
arising from the inner  disk regions presents a puzzle; the accretion
rates onto the star, coupled with  reasonable estimates of the disk
viscosity, suggest that the accreting material populating  the inner
disk must be highly depleted in small grains relative to the
populations of such  grains in an unprocessed `interstellar mix' of
gas and solid material. The mechanism(s)  needed to deplete the small
grain population are currently unknown (though see Najita \etal\ 2007
for a discussion of various possibilties).

Determining whether the formation of a Jovian mass planet  is a
plausible explanation  for the thin/thick disks observed among  the
intermediate mass stars in our sample  would require measurements of
(a) outer disk masses inferred from millimeter and sub-millimeter 
measurements, in order to place thin/thick objects in the $M_{\rm
acc}$ vs $M_{\rm disk}$ plane; (b) accretion rates inferred from
modeling Balmer or Brackett line  emission profiles (e.g., Muzerolle,
Calvet, \& Hartmann 1998; Muzerolle, Calvet, \&  Hartmann 2001); (c)
observation of the distribution of gas in the inner and outer disks 
via measurement of line profiles for gas tracers such as CO and
H$_2$O; such mid-IR  measurements can diagnose the presence/absence
of gaps (e.g., Najita \etal\ 2010;  Brittain, Najita \& Carr 2009;
Najita, Crockett, \& Carr 2008). If the planet formation  hypothesis
is correct, the thin/thick disks should exhibit low accretion rates
compared to  those typical of optically thick accretion disks of
comparable mass and CO or H$_2$O  profiles consistent with ``gaps''
in the distribution of gas consistent with the gas flow  models
proposed by, for example, Lubow \etal\ (1999). Unfortunately, with
the exception  of accretion rates estimated via Balmer or Brackett
line profiles, the sensitivity of current  ground-based optical/IR
and mm telescopes is insufficient to carry out such measurements  for
clusters as distant as IC 1805. However, searches for analogs of
thin/thick disks  among more proximate regions of intermediate mass
star formation (e.g., the North  American Nebula at a distance
$\sim$600 pc; see Guieu \etal\ 2009 and Rebull \etal\ 2010) may
provide a target  list of sufficient richness for more detailed
studies.

\subsection{SEDs consistent with empty inner holes and optically thin
outer disks (empty/thin)}

Such disks show no evidence of significant excess ($>$0.5 mag; see
\S\ref{sec:SED}) above the stellar photosphere for wavelengths
$\lambda < \lambda_{\rm initial}$, where  $\lambda_{\rm initial}$ is
the shortest wavelength at which significant excess emission is 
observed to lie above photospheric levels. These stars are nearly 10
times more  common than the optically thick accretion disk population
and comprise 8.9\% of the  total population of IC 1805 members with
$M>$2 \msun. 

Unfortunately, all but one of the stars in IC 1805 that fall in this
category lack 24 \mum\  measurements, owing primarily to a
combination of sensitivity and confusion between  source and nebular
background at this wavelength. That one star (No.\ 340) could be a 
variant within the thin/thick disk objects and thus could be in the
process of  forming extrasolar giant planets.  For the other stars,
we cannot distinguish between this  possibility or an alternative,
namely that they might be analogs of the objects discussed  by Currie
\etal\ (2009): objects in which the observed optically thin excess
emission at  8 \mum\ among our IC 1805 sample of empty/thin disks
derives directly  from the growth of planets following the precepts
of Kenyon and Bromley (2005; 2008),  i.e., the  ``collisional
cascade'' model.  In this case, the emission seen at 24 \mum\ even 
for the older stars studied by Currie \etal\ (see also Hernandez
\etal\ 2006) fall below the  upper limits measured at 24 \mum\ for
the IC 1805 stars in this category.

The Kenyon and Bromley simulations start with a swarm of $\sim0.1$-10
km icy bodies that  collide and rapidly grow ($t \sim$ 1-2 Myr) to a
size of $\sim$1000 km (runaway growth phase).  Once icy bodies grow to
1000 km sizes, they stir the leftover planetesimals to much  higher
velocities. These higher velocities reduce the rate of planet growth
(gravitational  focusing is less important), and the planets enter a
phase of oligarchic growth.  The  higher velocities characteristic of
the stirred planetesimal swarm also cause more  energetic collisions
between planetesimals, resulting in more fragmentation and an 
increase in the dust production rate. Currie \etal\ (2007; 2008)
present a series of color-color diagrams in which they track the
evolution of IRAC and MIPS-24 excesses as a  function of time.
Qualitatively, there is an initial rise in excess emission resulting
from  planetesimal formation and stirring, followed by a peak, and an
exponential decay  resulting from a diminishing number of remnant
planetesimals, combined with removal  of small dust by radiation
pressure and Poynting-Robertson drag. For their `canonical'  model,
the peak in 8 \mum\ emission vs.\ time occurs early, $t\sim$ 1 Myr,
and decays  rapidly thereafter, while the peak for 24 \mum\ occurs at
$t\sim$10 Myr, reflecting the more  leisurely timescales for planet
building in the outer disk (where lower Keplerian rotation  speeds
obtain) as compared with the inner disk. Hence, as stars evolve over
the age range  $t \sim$ 1-10 Myr, the emission at 8 \mum\ and
shortward is expected to drop rapidly, while  the excess at 24 microns
increases. Since our IC 1805 sample spans ages 1-3 Myr, it is tempting
to attribute the observed 8 micron excess emission to the stirring
of planetesimals in the inner disk. We note, however, that the
magnitude of the 8 \mum\ excess  emission seen in IC 1805, typically
slightly more than one magnitude, appears to  exceed that predicted by
the canonical Kenyon and Bromley model. Adjustments to their  model
(e.g., Currie \etal\ 2007) that posit a greater density of
planetesimals in the inner  regions and/or include the effects of PAH
emission from the dust might plausibly  produce the larger excesses
that we observe (see \S\ref{sec:SED}).

\subsection{SEDs consistent with empty inner and optically thick outer 
disks (empty/thick)}

The SEDs in this category exhibit flux densities consistent with
photospheric emission at  wavelengths 8 \mum\ and shortward, and 24
\mum\ flux densities consistent with an optically thick  disk. Only
four objects in our sample have SEDs consistent with ``empty/thick''
disks (that  is, slightly less than half of the number of optically
thick accretion disks). Two mechanisms could explain such disks: (a)
tidal isolation of the inner and outer disk  by a supra-Jovian mass
($>3~M_J$) planet, a brown dwarf, or a stellar companion; or (b) 
isolation of the inner and outer disk via photoevaporation driven by 
EUV radiation from the central star. In particular, isolation, and 
rapid inner disk clearing, occurs when the photoevaporation rate
driven by the central  star exceeds the rate at which material from
the outer disk accretes inward to the central  star, thus precluding
inward accretion. Once this happens, material in the inner disk is 
isolated from the outer disk, following which it accretes rapidly
onto the star (on a  viscous timescale). After the disk develops an
empty ``inner hole,'' radiation from the  central star can illuminate
the inner disk ``wall'' located at an initial radius, $R_{\rm hole}$,
which is  equal to the radius at which the escape speed of
radiatively heated disk material  exceeds the gravitational pull of
the star-disk system, $R_{\rm evap}$. Once the inner disk  empties,
photoevaporation begins to erode the disk not only from the surface,
but from  ``inside out'' as $R_{\rm hole}$ increases in response to
photoevaporation of material from the  inner wall of the disk; the
demise of the disk is near!

The factors (e.g., Alexander \etal\ 2006; Clarke, Gendrin, \&
Sotomayor 2001) that affect  when and whether this latter mechanism
becomes important depend on (i) the  frequency distribution of initial
disk accretion rates, which in turn reflect the distribution  of disk
sizes and masses; (ii) the decrease in disk accretion rate with time;
and (iii) the photoevaporation rate. Disks  with lower initial accretion
rates (perhaps those with lower initial masses) will reach the 
``balance point'' between photoevaporation and accretion rates earlier
than their brethren  among higher initial accretion rate (higher mass)
disks. Whether photoevaporative  clearing ``wins'' in any particular
disk depends on both initial conditions and the  (unknown, but
eventually constrainable) rates at which giant planet-building takes
place. 

From the discussion above, it appears plausible to assume that
photoevaporation plays  a major role in driving the evolution of disks
among stars more massive than 2 \msun.  Recall that the overwhelming
majority of intermediate (2-4 \msun) mass stars in IC 1805  have
already lost their disks at very early ages ($t <$ 2 Myr), and
moreover, higher mass  stars (which are expected to 
photoevaporatively erode  their disks on timescales $t \sim$ 0.5  Myr)
are observed to lack any remnant accretion disks. Both observations
are  qualitatively consistent with photoevaporation models (Gorti,
Dullemond, \&  Hollenbach 2009). Are the empty/thick disks among our
sample evidence of the ongoing  effects of photoevaporation on the few
surviving optically thick accretion disks? 

Two factors provide some, albeit relatively weak, support for this
proposal. First, the  radius ($R_{\rm evap}$) at which disks become
susceptible to photoevaporation  is $\sim$20 AU for a  3 \msun\ star
(Gorti, Dullemond, \& Hollenbach 2009). As noted above, isolated 
material within $R_{\rm evap}$ accretes onto the star rapidly (on a
timescale $\sim10^5$ years),  leaving an inner disk hole of initial
dimension $R_{\rm hole} \sim R_{\rm evap}$. For a disk with $R_{\rm
hole} \sim 20$  AU, the observed SED should exhibit no excess
shortward of 8 \mum, and excess  emission consistent with that from
an optically thick disk at 24 \mum. The observed  morphology of the
SEDs observed for our empty/thick disks jibes with SEDs expected  for
disks whose structure is determined by ongoing photoevaporation. The
lifetime of  empty inner/thick outer disk systems is short: several
$\times 10^5$ years. Hence we would  expect that, were
photoevaporation responsible for producing the majority of
empty/thick  systems, such systems should comprise about 10-20\% of
the number of optically thick  accretion disk systems that survive
for $\sim$2 Myr (the age of most stars in IC 1805). The  observed
fraction, 4/8, is larger but perhaps not inconsistent with this 
estimate, given the small numbers of empty/thick and optically thick
accretion disk systems.

More robustly choosing between the hypotheses that companions or
photoevaporation  are the predominant cause of this rare disk type
requires (a) sensitive searches for  companions; and (b) measurements
aimed at constraining accretion rates, disk gas mass, and outer disk
masses. If photoevaporation is producing empty/thick disks, then  we
expect to see no evidence of accretion, no evidence of inner disk gas,
and low outer  disk masses (consistent with the low accretion rates
necessary if photoevaporation is to  isolate outer from inner disks).
If companions produce the inner holes, then we should  be able to
detect spectral features from brown dwarf and cool stellar companions 
directly via high signal-to-noise  spectroscopic searches (e.g., Prato
\etal\ 2002). If supra-Jovian mass planets are  responsible, it will
be necessary to carry out high contrast ratio adaptive optics imaging
of nearby  analogs of the thin/thick objects in IC 1805 (e.g., those
in IC 348; Muzerolle \etal\ 2006). 

\section{Conclusions}
\label{sec:concl}

\subsection{Main Conclusions}

We have carried out a study of the rich, star-forming complex IC 1805
based on optical, NIR, and Spitzer mid-IR photometry, and
classification resolution optical  spectroscopy. Our basic conclusions
are as follows.

1.	The stars more massive than $M \sim$ 12 \msun\ appear to be
concentrated within a  relatively small region of the parent
molecular cloud for the IC 1805 complex. The  ages of these stars
were estimated to lie between 1-3 Myr by Massey \etal\ (1995a). 
Given the uncertainty in estimating the ages of intermediate mass
pre-main-sequence stars, we have assumed that the ages of the 2-4
\msun\ stars  studied in this paper are also 1-3 Myr.

2.	We find that (i) no stars more massive than $M \sim 4$ \msun\
exhibit optically thick IR emission, as is characteristic of the
circumstellar material surrounding Herbig AeBe stars; 
(ii) among stars in the mass range $2 < M/M_{\odot} < 4$,
less than 2\% of the stars exhibit such excesses.

3.	Examination of the SEDs for those stars showing IR excesses
reveals four additional distinct categories: (a) emission at $H$-band
and longer wavelengths that lies significantly below that expected
for an optically thick, geometrically flat reprocessing disk; we
denote these as ``optically thin'' disks; (b) excess emission
smaller  than that expected for optically thick accretion disks at
$JHK_s$ and, in the case of  No.\ 882, also in the IRAC bands, but
with excess emission at 24 \mum\  consistent with that expected for
optically thick accretion disks; we call these thin  inner
region/thick outer region disks (``thin/thick''); (c) no excess
emission significantly above  photospheric levels at $\lambda <
\lambda_{\rm initial}$, and excess emission at least 2.5   times
smaller than that expected for optically thick accretion disks at
$\lambda =  \lambda_{\rm initial}$; these disks are denoted empty
inner region/thin outer region (``empty/thin''); and  (d) no excess
emission above photospheric levels through the IRAC bands, and 
excess emission consistent with that expected for an optically thick
accretion disk  at  24 \mum; we call these empty inner region/thick
outer region disks (``empty/thick'').  Among  stars with masses
greater than 2 \msun, the fraction of the total population of IC 1805
members in each of these categories is: (optically thick [AeBe])
1.3\%; (optically thin) 0.9\%; (thin/thick) 0.4\%; (empty/thin) 
8.2\%; and (empty/thick) 0.7\% 

4.	We interpret stars in the ``optically thick'' category as the
tail end of a population of optically thick  accretion disks that
have survived for $t \sim$ 3 Myr; 98\% of their brethren have 
already transitioned from the accretion phase to the other categories
described above, or  show no evidence of remnant or second generation
``debris'' disk material  (at least as yet).  We  estimate that the
median lifetime for optically thick accretion disks in this mass 
range is $t \sim$ 0.5 Myr, and speculate that the 2\% remaining in
the accretion phase  until $t \sim$ 2 Myr are those that had initial
disk masses near the high end of the  distribution of disk masses
that characterized intermediate mass stars on the  birthline (just
after they detach from their infalling envelopes).

5.	 We suggest that photoevaporation, driven by the powerful
extreme and far  ultraviolet radiation fields characteristic of stars
with $M > 4$ \msun\ likely accounts  for the fact that none of the
high-mass stars in this survey shows evidence of excess emission
arising  in an optically thick accretion disk; such disks are
destroyed by photoevaporation  on timescales $t <<$ 0.5 Myr. We argue
that photoevaporation as well represents  an attractive mechanism for
explaining the rapid destruction of disks among stars  with $2 <
M/M_{\odot} < 4$.

6.	We propose that stars in the ``optically thin'' category  are
either surrounded  by gas-dominated disks, analogous to the decretion
disks believed to surround  classical, rapidly rotating Be stars, or
by debris disks in which vigorous dust  production is ongoing both in
the terrestrial and outer planet regions.

7.	 We suggest that the ``thin inner disk/ thick outer disk'' 
SED likely results from the formation of a Jovian mass planet that has
produced  a gap inward of the orbital radius of the planet. Such disks
find their analog  among well-studied ``transition disks'' around low
mass stars that share this SED  morphology with the intermediate mass
stars studied here.

8.	 We suggest that stars in the fourth category owe their
``empty inner disk/thin outer disk''  SEDs to the production of dust
by a planetesimal swarm located at disk radii $r >$  10 AU.

9.	 Stars in the last category, those with empty inner disks and
optically thick outer disks (``empty/thick''), seem best explained by
positing either a companion of mass sufficient to tidally  isolate
material located in the outer optically thick disk; or a disk in
which the  accretion rate has dropped below the photoevaporation
rate. Disks in this latter  state quickly develop an inner hole, as
material inward of the photoevaporation  radius ($R_{\rm evap}$)
accretes onto the central star on a viscous timescale, while 
material in the outer disk is prevented from accreting inward from
$R_{\rm evap}$.

\subsection{Implications for Planet Building}

Optically thin disks, if they are indeed building planets, are
likely  forming planetesimals and/or low mass planets. If so, then the
observed optically thin  emission arises from dust produced in
collisions between planetesimals spanning a  wide range ($<$1 to
$\sim$20 AU). However, before accepting this explanation, it is
essential to  establish that the observed excesses arise primarily
from dust rather than gas.

Thin/thick disks are the most likely candidates to be forming giant
planets.  Of  a sample of nearly 500 stars of intermediate mass 
($M \sim$ 2-4 \msun), only two  objects exhibit this type of SED. If
these objects are indeed forming Jovian mass planets, it  is
interesting to note that the orbital distance of the planet is
$\sim$20 AU (that is, the size of  the `gap' inward of the region of
the disk which produces emission consistent with that  observed for
an optically thick disk at 24 \mum). This distance roughly
corresponds to  the ice sublimation radius for a disk surrounding a 3
\msun\ star, the radius that many  theories of giant planet formation
posit as a natural place for forming giant planets, where a
`snowstorm' at the ice sublimation radius may accelerate the rapid
formation of  a massive solid core that ultimately accretes
surrounding gas (Ida \& Lin 2008).

The 45 objects found in the empty/thin category may be in a very early
stage of planetesimal/planet  building in which planetesimal building
in the inner disk ($r <$ 10 AU) is already well  underway. 

Even if all of the objects in the optically thin, thin/thick, and
empty/thin categories are in the process of forming either 
planetesimals or planets, not more than 10\% of the stars in IC 1805
show evidence of  such activity.  This is far below the fraction in
older groups (e.g., Hernandez \etal\ 2006).   The absence of
significant numbers of stars exhibiting signatures of nascent
planet-building may well result from our inability to measure 24 \mum\
flux densities (as opposed to upper limits or the star being off the
edge of the map) for all but 5\% (48 of the 974) of the stars in the
total sample. If planetesimal building mimics the `canonical' model
first suggested by Kenyon \&  Bromley (2005; 2008), then we might
expect for most stars with ages 1-2 Myr and older, the  `wave' of
planet building has already passed through the inner portions of the
disk  observed in the IRAC bands. If so, then planetesimal collisions
(and resulting production  of dust debris) should be restricted to the
outer disk, in which case IR excess emission  will be observable only
at wavelengths 24 \mum\ and beyond. The significant fraction  (46\%)
of disks in Orion Ia ($t\sim$10 Myr) and in Orion Ib ($t \sim$ 5 Myr;
38\% disks) that show  optically thin emission at 24 \mum, but exhibit
no excess emission at shorter  wavelengths (Hernandez \etal\ 2006)
would suggest that we could indeed, be missing a  large number of such
disks in IC 1805. Observations of a sample of intermediate mass  stars
in the North American Nebula ($d \sim$ 600 pc), where more robust 24
\mum\  measurements are possible, would provide a test of the
hypothesis that a significant  number of young ($t <$ 2 Myr)
intermediate mass stars are undergoing planetesimal  collisions in
their outer disks (following a short-lived phase of such
planet-building in  their inner disks).

\acknowledgements 

The authors wish to thank Diane Harmer and Darryl Willmarth for making
most of the spectroscopic observations at WIYN.  We also thank  Lori
Allen,  Uma Gorti, Lee Hartmann, David Hollenbach,  Xavier Koening,
Greg Laughlin, Doug Lin, and Joan Najita for a variety of stimulating 
discussions, insightful comments and important critical remarks. We
thank Jane Greaves for her generosity in offering access to her MIPS
data during the early stages of this investigation. This work is based
in  part on observations made with the Spitzer Space Telescope, which
is operated by the  Jet Propulsion Laboratory, California Institute of
Technology under a contract with  NASA.  Support for this work was
provided by NASA through an award  issued by JPL/Caltech.

This research has made use of NASA's Astrophysics Data System (ADS)
Abstract Service, and of the SIMBAD database, operated at CDS,
Strasbourg, France.  This research has made use of data products from
the Two Micron All-Sky Survey (2MASS), which is a joint project of the
University of Massachusetts and the Infrared Processing and Analysis
Center, funded by the National Aeronautics and Space Administration
and the National Science Foundation.  These data were served by the
NASA/IPAC Infrared Science Archive, which is operated by the Jet
Propulsion Laboratory, California Institute of Technology, under
contract with the National Aeronautics and Space Administration.  This
research has made use of the Digitized Sky Surveys, which were
produced at the Space Telescope Science Institute under U.S.
Government grant NAG W-2166. The images of these surveys are based on
photographic data obtained using the Oschin Schmidt Telescope on
Palomar Mountain and the UK Schmidt Telescope. The plates were
processed into the present compressed digital form with the permission
of these institutions.

The research described in this paper was partially carried out at the
Jet Propulsion Laboratory, California Institute of Technology, under
contract with the National Aeronautics and Space Administration.

\appendix
\section{Complete set of SEDs}

\begin{figure*}[tbp]
\epsscale{0.75}
\plotone{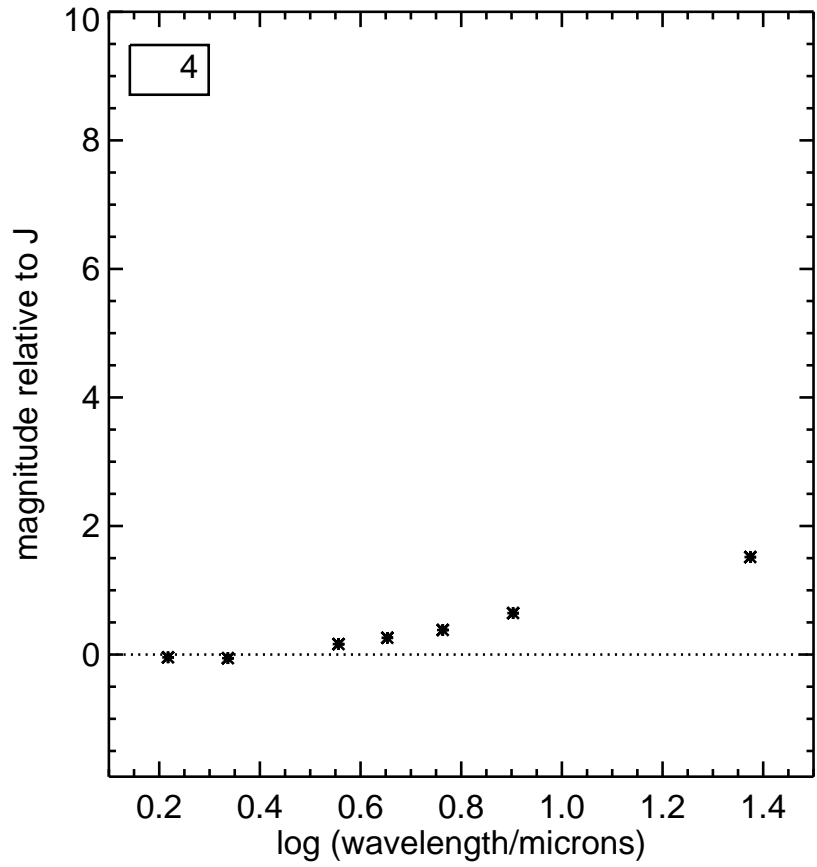}
\caption{Magnitudes relative to $J$ for stars with excesses, listed in
order of optical number.  }
\label{fig:appendixpostagestamps}
\end{figure*}

Figure~\ref{fig:appendixpostagestamps}, available online only, shows
the complete set of SEDs for the stars with excesses.   The magnitudes
relative to $J$ have been corrected for reddening, and the
photospheric emission has been subtracted.  

\begin{thebibliography}{}

\bibitem{} Alexander, R.D., Clarke, C. J., \& Pringle, J. E. 2006, \mnras, 369, 229
\bibitem{}  Andrews, S. M., \& Williams, J. P. 2005, \apj, 631, 1134
\bibitem{}  Andrews, S. M., \& Williams, J. P. 2007, \apj, 659, 705
\bibitem{} Brittain, S. D., Najita, J. R., Carr, J. S. 2009, \apj 702, 85
\bibitem{} Calvet, N., et al. 2005, \apj, 630, L185  
\bibitem{} Calvet, N., Hartmann, L. \& Strom, S.E. 2000 in Protostars and Planets IV 
(University of Arizona Press, Tucson) p 377.
\bibitem{} Cieza, L. A., et al. 2010, \apj, 712, 925
\bibitem{} Clarke, C. J., Gendrin, A., \& Sotomayor, M. 2001, \mnras, 328, 485
\bibitem{} Cote, J. \& Waters, L. B. F. M. 1987, A\&A 176, 93
\bibitem{} Crawford, D.L., \& Perry, C. L. 1976, \aj, 81, 419
\bibitem{} Currie, T., et al. 2007, \apj, 663, L105
\bibitem{} Currie, T., et al. 2008, \apj, 672, 558
\bibitem{} Currie, T., \& Kenyon, S. 2009, \aj, 138, 703
\bibitem{} Currie, T., et al. 2009, \apj, 698, 1
\bibitem{} Dahm, S. E., \& Hillenbrand, L. A. 2007, \aj, 133, 2072
\bibitem{} Dougherty, S.M., Taylor, A. R., \& Clark, T. A. 1991, \aj, 102, 1753
\bibitem{} Dougherty, S.M. et al. 1994, A\&A 290, 609
\bibitem[engel]{engel}Engelbracht, C., \etal, 2007, \pasp, 119, 994
\bibitem{} Evans, N., et al. 2009, arXiv:0901.1691v1
\bibitem[fazio]{fazio}Fazio, G., \etal 2004a, ApJS, 154, 10
\bibitem{} Fazio, G. G., et al. 2004b, \apjs, 154, 39
\bibitem{}Finkenzeller, U. 1985, \aa, 151, 340
\bibitem{} Flaherty, K. M., et al. 2007, \apj, 663, 1069
\bibitem[gordon]{gordon}Gordon, K., \etal 2005, \pasp, 117, 503
\bibitem{} Guieu, S., et al. 2009, \apj, 697, 787
\bibitem{} Gorti, U., Dullemond, C. P., \& Hollenbach, D. 2009, \apj, 705, 1237
\bibitem{} Haisch, K. E., Jr., Lada, E. A., \& Lada, C. J., 2001, \apj, 553, L153
\bibitem{} Hernandez, J., et al. 2005, \aj, 129, 856
\bibitem{} Hernandez, J., et al. 2006, \apj, 652, 472
\bibitem{} Hernandez, J., et al. 2007, 662, 1067
\bibitem{} Hillenbrand, L. A. 1997, \aj, 113, 1733
\bibitem{} Hillenbrand, L. A., Strom, S. E., Vrba, F. J., \& Keene, J. 1992, \apj, 397, 613
\bibitem{} Hillenbrand, L. A., Massey, P., Strom, S. E., \& Merrill, K. M. 1993, \aj, 106, 
1906
\bibitem{} Ida, S., \& Lin, D. N. C. 2008, \apj, 685, 584 
\bibitem{} Indebetouw, R., et al. 2005, \apj, 619, 931
\bibitem{} Kang, M., Bieging, J. H., Povich, M. S., \& Lee, Y. 2009, \apj, 706, 83
\bibitem{} Kenyon, S.J., \& Bromley, B. C., 2005, \aj, 130, 269
\bibitem{} Kenyon, S.J., \& Bromley, B.C., 2008, \apjs, 179, 451
\bibitem{} Koenig, X., \& Allen, L. A. 2010, in preparation
\bibitem{} Lada, C. J., et al. 2006, \aj, 131, 1574L
\bibitem{} Low, F.J., et al. 2005, \apj, 631, 1170
\bibitem{} Lubow, S.H., Seibert, M., \& Artymowicz, P. 1999, \apj, 526, 1001
\bibitem[mako]{mako}Makovoz, D., \& Marleau, F. 2005, \pasp, 117, 1113
\bibitem{} Massey, P., Johnson, K. E., \& Degioia-Eastwood, K. 1995a, \apj, 454, 151
\bibitem{} Massey, P., Lang, C. C., Degioia-Eastwood, K., \& Garmany, C. D. 1995b, 
\apj, 438, 188
\bibitem{} Massey, P., Puls, J., Pauldrach, A. W. A., Bresolin, F., Kudritzki, R. P., \& Simon, T. 2005, \apj, 627, 477
\bibitem{} Merin, B., \etal\ 2010, \apj, 718, 1200
\bibitem{} Muzerolle, J., Calvet, N., \& Hartmann, L. 1998, \apj, 492, 743
\bibitem{} Muzerolle, J., Calvet, N., \& Hartmann, L. 2001, \apj, 550, 944
\bibitem{} Muzerolle, J., et al. 2003, \apj, 592, 266
\bibitem{} Muzerolle, J. et al. 2006, \apj 643, 1003
\bibitem{}Muzerolle, J., Allen, L. E., Megeath, S. T., Hernandez, J., \& Guthermuth, R. A. 
2010, \apj, 708, 1107
\bibitem{} Najita, J.R., Strom, S. E., \& Muzerolle, J. 2007, \mnras, 378, 369
\bibitem{} Najita, J.R., Crockett, N. \& Carr, J. 2008, \apj, 687, 1168
\bibitem{} Najita, J.R., Carr, J. S., Strom, S. E., Watson, D.M., Pascucci, I.,  Hollenbach, 
D., Gorti, U. \& Keller, L. 2010, \apj, 712, 274
\bibitem{} Prato, L, Simon, M., Mazeh, T., McLean, I. S., Norman, D., and Zucker, S. 
2002, \apj, 569, 863
\bibitem[bpmg]{bpmg}Rebull, L., \etal, 2008, ApJ, 681, 1484
\bibitem{} Rebull, L., \etal 2010, ApJ, in preparation
\bibitem{} Rieke, G. H., \& Lebofsky, M. J. 1985, \apj, 288, 618
\bibitem[mips]{mips}Rieke, G., \etal 2004, ApJS, 154, 25
\bibitem{} Robitaille, T. P., Whitney, B. A., Indebetouw, R. , Wood, K., \& Denzmore, P. 
2006, \apjs, 167, 256
\bibitem{} Sagar, R., Myakutin, V. I., Piskunov, A. E., \& Dluzhnevskaya, O. B. 1988, 
\mnras, 234, 831
\bibitem{} Schaller, G., Schaerer, D., Meynet, G., \& Maeder, A. 1992, A\&AS, 96, 269
\bibitem{} Siess L., Dufour E., Forestini M. 2000, A\&A, 358, 593 
\bibitem{} Skrutskie, M. F., Dutkevitch, D., Strom, S. E., Edwards, S., Strom, K. M., \& 
Shure, M. A. 1990,  \aj, 99, 1187
\bibitem[2mass]{2mass}Skrutskie, M., \etal 2006, AJ, 131, 1163
\bibitem{} Strom, S. E. 1972, \pasp, 84, 745
\bibitem{} Strom, S. E., Strom, K. M., Brooke, A. L., Bregman, J., \& Yost, J. 1972, \apj, 
171, 267
\bibitem{} Strom, K. M., Strom, S. E., Edwards, S., Cabrit, S., \& Skrutskie, M. F. 1989, 
\aj, 97, 1451
\bibitem{} Testi, L., Palla, F. and Natta, A. 1998 A\&AS 133, 81 
\bibitem{} Vasilevskis, S., Sanders, W.L., \& van Altena, W. F. 1965 \aj , 70, 806
\bibitem{} Warren, P. H., \& Hesser, J. E. 1978, \apjs, 36, 497
\bibitem[werner]{werner}Werner, M., \etal, 2004, ApJS, 154, 1



\end{thebibliography}
\end{document}